\documentclass{article}

\usepackage{microtype}
\usepackage{graphicx}
\usepackage{subfigure}
\usepackage{booktabs}
\usepackage{hyperref}
\usepackage{placeins}
\usepackage{xurl}
\usepackage[subtle]{savetrees}

\usepackage[accepted]{mlsys2025}

\usepackage{xspace}
\usepackage{etoolbox}
\usepackage{comment}
\usepackage{amsmath}
\usepackage{amssymb}
\usepackage{enumitem}
\usepackage{multirow}
\usepackage{tabularx}
\usepackage{xcolor}
\usepackage[nameinlink]{cleveref}

\newcommand{\sys}{\textsc{ShadowPrefill}\xspace}          %

\definecolor{ForestGreen}{RGB}{34,139,34}
\newtoggle{showmarks}
\toggletrue{showmarks}             %

\iftoggle{showmarks}{
  \newcommand{\todo}[1]{{\color{red}\textbf{[TODO: #1]}}}
  \newcommand{\authornote}[3]{\textcolor{#1}{~[#2: #3]~}}
}{
  \newcommand{\todo}[1]{\unskip}
  
  \newcommand{\authornote}[3]{\unskip}
}

\newcommand{\vs}{\emph{vs.}\xspace}
\newcommand{\myparagraph}[1]{\vspace{\smallskipamount}\noindent\textbf{#1.\xspace}}

\newcommand{\myparagraphnodot}[1]{\vspace{\smallskipamount}\noindent\textbf{#1\xspace}}

\newcounter{insightcounter}

\usepackage{xcolor}
\definecolor{darkgreen}{RGB}{0,90,50}
\newcommand{\nosh}{\textcolor{darkgreen}{{\scalebox{0.9}{\textsf{No Shadow}}}}\xspace}
\newcommand{\hosttier}{\textcolor{darkgreen}{\scalebox{0.9}{\textsf{Host Tier}}}\xspace}
\newcommand{\harvest}{\textcolor{darkgreen}{\scalebox{0.9}{\textsf{Harvest}}}\xspace}
\newcommand{\harvestslo}{\textcolor{darkgreen}{\scalebox{0.9}{\textsf{Harvest-SLO}}}\xspace}
\newcommand{\full}{\textcolor{darkgreen}{\scalebox{0.9}{\textsf{Harvest-SLO+Host}}}\xspace}
\newcommand{\predictive}{\textcolor{darkgreen}{\scalebox{0.9}{\textsf{Predictive}}}\xspace}
\newcommand{\dedicated}{\textcolor{darkgreen}{\scalebox{0.9}{\textsf{Dedicated}}}\xspace}
\newcommand{\eager}{\textcolor{darkgreen}{\scalebox{0.9}{\textsf{Eager}}}\xspace}
\newcommand{\oracle}{\textcolor{darkgreen}{\scalebox{0.9}{\textsf{Oracle}}}\xspace}
\newcommand{\oraclehost}{\textcolor{darkgreen}{\scalebox{0.9}{\textsf{Oracle+Host}}}\xspace}
\newcommand{\alwayscheap}{\scalebox{0.9}{\textsf{Always Cheap}}\xspace}
\newcommand{\alwaysexp}{\scalebox{0.9}{\textsf{Always Expensive}}\xspace}

\newcommand{\Smodel}{$S$\xspace}
\newcommand{\Lmodel}{$L$\xspace}

\newcommand{\nAccCostCutMax}{34}
\newcommand{\nAccCostCutMin}{22}
\newcommand{\nAccGainLargeMax}{13.6}
\newcommand{\nAccGainLargeMin}{6.6}

\newcommand{\nAllCostFull}{29.5}
\newcommand{\nAllCostFullMax}{46.1}
\newcommand{\nAllCostFullMin}{4.3}

\newcommand{\nAllLatFull}{6.3}
\newcommand{\nAllLatFullMax}{17.7}
\newcommand{\nAllLatFullMin}{1.3}

\newcommand{\nAllSwFull}{11.4}
\newcommand{\nAllSwFullMax}{21.4}
\newcommand{\nAllSwFullMin}{3.5}

\newcommand{\nAllTaskFullMax}{17.7}

\newcommand{\nAllTtftFull}{44.9}
\newcommand{\nAllTtftFullMax}{56.4}
\newcommand{\nAllTtftFullMin}{9.2}

\newcommand{\nAppBgFullBurst}{6.5}
\newcommand{\nAppBgFullCode}{5.9}
\newcommand{\nAppBgSloAttBurst}{0.4}
\newcommand{\nAppBgSloBurst}{5.7}
\newcommand{\nAppBgSloCode}{5.1}
\newcommand{\nAppBgSloConv}{3.3}
\newcommand{\nAppBwSwFullMax}{13.0}
\newcommand{\nAppBwSwFullMin}{12.0}
\newcommand{\nAppCapFullOne}{6.0}
\newcommand{\nAppCapFullQ}{1.6}
\newcommand{\nAppCapSloOne}{4.2}
\newcommand{\nAppCapSloQ}{-0.2}
\newcommand{\nAppChunkBigFullMax}{5.1}
\newcommand{\nAppChunkBigFullMin}{4.9}

\newcommand{\nAppChunkFullSmall}{5.9}
\newcommand{\nAppCtrlAttMax}{0.27}
\newcommand{\nAppCtrlAttMin}{-0.04}
\newcommand{\nAppDeltaFullTen}{4.0}
\newcommand{\nAppDeltaFullZero}{5.4}
\newcommand{\nAppDeltaSloMax}{3.6}

\newcommand{\nAppHostSwFullDefault}{12.9}
\newcommand{\nAppHostSwFullSmall}{11.9}
\newcommand{\nAppHostSwOr}{23.0}
\newcommand{\nAppHostSwTierDefault}{2.4}
\newcommand{\nAppHostSwTierSmall}{1.3}
\newcommand{\nAppInstFullSixteen}{6.2}
\newcommand{\nAppInstFullTwo}{5.3}
\newcommand{\nAppInstOrSixteen}{11.7}
\newcommand{\nAppInstOrTwo}{9.6}
\newcommand{\nAppMarginFullMax}{5.1}
\newcommand{\nAppMarginFullMin}{4.9}
\newcommand{\nAppMinFullFiveTwelve}{2.8}
\newcommand{\nAppMinFullKilo}{1.4}
\newcommand{\nAppMinFullSixtyFour}{6.6}

\newcommand{\nAppNmAllFive}{6.3}
\newcommand{\nAppNmAllThree}{6.3}
\newcommand{\nAppNmEnergyAllFive}{2.8}
\newcommand{\nAppNmEnergyTopFive}{0.1}
\newcommand{\nAppNmOrFive}{14.6}
\newcommand{\nAppNmOrTwo}{10.0}
\newcommand{\nAppNmSwAllFive}{10.1}
\newcommand{\nAppNmSwAllTwo}{12.9}
\newcommand{\nAppNmSwSessFive}{5.1}
\newcommand{\nAppNmTopFive}{4.6}
\newcommand{\nAppNmTopThree}{4.2}
\newcommand{\nAppNmWasteAllFive}{89.0}
\newcommand{\nAppNmWasteAllTwo}{68.6}
\newcommand{\nAppNmWasteTopFive}{73.0}

\newcommand{\nAppPdOrOne}{11.7}
\newcommand{\nAppPdOrThree}{7.3}
\newcommand{\nAppPdSloOne}{8.7}
\newcommand{\nAppPdSloThree}{4.9}
\newcommand{\nAppPolFullAcc}{5.5}
\newcommand{\nAppPolFullCost}{5.0}
\newcommand{\nAppPolFullRand}{5.7}

\newcommand{\nAppPolSwSessRand}{4.9}
\newcommand{\nAppRtFullHigh}{5.8}
\newcommand{\nAppRtFullLow}{-0.2}
\newcommand{\nAppRtOrHigh}{10.9}
\newcommand{\nAppRtOrLow}{4.1}
\newcommand{\nAppRtSwFullHigh}{10.8}
\newcommand{\nAppRtSwFullLow}{13.4}
\newcommand{\nAppRtSwOrHigh}{19.0}
\newcommand{\nAppRtSwOrLow}{30.8}
\newcommand{\nAppSwSessData}{3.2}
\newcommand{\nAppSwSessHigh}{5.0}
\newcommand{\nAppSwSessLow}{0.5}
\newcommand{\nAppTpFull}{3.5}
\newcommand{\nAppTpOr}{8.0}
\newcommand{\nAppTpOrDefault}{10.0}
\newcommand{\nAppTpSlo}{2.9}

\newcommand{\nAppXferSloMax}{5.4}
\newcommand{\nAppXferSloMin}{4.9}

\newcommand{\nBenchCostFull}{28.5}

\newcommand{\nBenchHitFull}{82.1}

\newcommand{\nBenchHitNone}{75.0}

\newcommand{\nBenchHitOracleHost}{84.4}

\newcommand{\nBenchLatCapture}{52}

\newcommand{\nBenchLatEagerUp}{7.0}
\newcommand{\nBenchLatFull}{5.0}

\newcommand{\nBenchLatHostTier}{0.4}
\newcommand{\nBenchLatOracle}{9.6}

\newcommand{\nBenchSwCapture}{54}

\newcommand{\nBenchSwDedicated}{18.0}
\newcommand{\nBenchSwFull}{10.4}
\newcommand{\nBenchSwHarvest}{6.9}

\newcommand{\nBenchSwOracle}{19.3}

\newcommand{\nBenchSwPredictive}{3.2}

\newcommand{\nBenchSwSlo}{8.8}

\newcommand{\nBenchTaskFull}{4.9}

\newcommand{\nBenchTtftFull}{44.1}

\newcommand{\nBgSloFull}{0.40}

\newcommand{\nBgSloSlo}{0.10}
\newcommand{\nColoEnergy}{0.8}
\newcommand{\nColoLat}{3.7}
\newcommand{\nColoSw}{7.7}
\newcommand{\nCostVsLargeMax}{56}
\newcommand{\nCostVsLargeMin}{49}

\newcommand{\nDedicatedGpus}{25}
\newcommand{\nEnergyDedicated}{11.7}
\newcommand{\nEnergyFull}{0.4}

\newcommand{\nEnergySlo}{0.8}

\newcommand{\nFbHitNone}{76.6}

\newcommand{\nFbHitSwitch}{52}

\newcommand{\nFbPromptMean}{13}
\newcommand{\nFbPromptPnn}{58}

\newcommand{\nFbSwitchRate}{26}

\newcommand{\nFbSwitchesPerSess}{3.1}

\newcommand{\nGhAccGainLarge}{13.6}

\newcommand{\nGhCostVsLarge}{54}

\newcommand{\nGhHitNone}{55.4}

\newcommand{\nGhHitSwitch}{44}

\newcommand{\nGhPromptMean}{63}
\newcommand{\nGhPromptPnn}{131}

\newcommand{\nGhSwitchRate}{26}
\newcommand{\nGhSwitchSlowdown}{1.5}
\newcommand{\nGhSwitchesPerSess}{2.0}

\newcommand{\nGhcpCostFull}{32.0}

\newcommand{\nGhcpCostHostTier}{17.8}

\newcommand{\nGhcpHitFull}{76.3}

\newcommand{\nGhcpHitNone}{55.4}

\newcommand{\nGhcpHitOracleHost}{81.3}

\newcommand{\nGhcpLatEagerUp}{3.2}
\newcommand{\nGhcpLatFull}{9.7}

\newcommand{\nGhcpLatHostTier}{8.1}

\newcommand{\nGhcpLatOracleHost}{13.8}

\newcommand{\nGhcpSwAbsNone}{19.0}
\newcommand{\nGhcpSwCapture}{52}

\newcommand{\nGhcpSwFull}{14.3}

\newcommand{\nGhcpSwOracleHost}{27.4}

\newcommand{\nGhcpSwSaveFull}{3.3}

\newcommand{\nGhcpSwSlo}{4.3}

\newcommand{\nGhcpTaskFull}{10.0}

\newcommand{\nGhcpTtftFull}{47.0}

\newcommand{\nGhcpTtftHostTier}{27.5}

\newcommand{\nHeadBurstPeak}{79}

\newcommand{\nHeadConvPeak}{69}

\newcommand{\nHitSameMax}{89}
\newcommand{\nHitSameMin}{80}
\newcommand{\nHitSwitchMax}{65}
\newcommand{\nHitSwitchMin}{44}

\newcommand{\nIdleCodeMed}{72}
\newcommand{\nIdleCodePeak}{8}
\newcommand{\nIdleCodeTrough}{85}

\newcommand{\nKnobAgeFullLong}{2.8}
\newcommand{\nKnobAgeFullTen}{5.2}

\newcommand{\nKnobAgeFullTwo}{4.8}
\newcommand{\nKnobAgeFullZero}{3.6}
\newcommand{\nKnobAgeHitDefault}{72}
\newcommand{\nKnobAgeHitLong}{54}

\newcommand{\nKnobAgeZeroTok}{2.1}
\newcommand{\nKnobCapFullDefault}{5.0}
\newcommand{\nKnobCapFullFive}{4.6}
\newcommand{\nKnobCapFullFull}{5.0}

\newcommand{\nKnobCapSloDefault}{3.3}

\newcommand{\nKnobShadowOcc}{1.7}

\newcommand{\nNoneVsLargeMax}{44}
\newcommand{\nNoneVsLargeMin}{28}
\newcommand{\nPdCapture}{78}
\newcommand{\nPdEnergy}{11.5}
\newcommand{\nPdLat}{5.2}

\newcommand{\nPdSw}{14.0}

\newcommand{\nRrHitNone}{69.2}

\newcommand{\nRrHitSwitch}{46}

\newcommand{\nRrPromptMean}{10}
\newcommand{\nRrPromptPnn}{45}

\newcommand{\nRrSwitchRate}{33}

\newcommand{\nRrSwitchesPerSess}{4.1}

\newcommand{\nSensAFullFirst}{0.7}
\newcommand{\nSensAFullLast}{20.5}

\newcommand{\nSensAOracleHostLast}{43.7}

\newcommand{\nSensBFullDefault}{6.7}
\newcommand{\nSensBFullFirst}{2.7}
\newcommand{\nSensBFullLast}{8.7}

\newcommand{\nSensCFullLast}{15.2}

\newcommand{\nSensCSloFirst}{4.4}
\newcommand{\nSensCSloLast}{0.7}

\newcommand{\nShadowWasteFull}{63}

\newcommand{\nSizeBigCost}{24}
\newcommand{\nSizeBigLat}{3}

\newcommand{\nSizeHitGainHi}{28}
\newcommand{\nSizeHitGainLo}{24}
\newcommand{\nSizeHugeHitGain}{0.2}
\newcommand{\nSizeMidCost}{37}
\newcommand{\nSizeMidLat}{4}
\newcommand{\nSizeOutHuge}{355}

\newcommand{\nSizePairs}{283}
\newcommand{\nSizeSmallCost}{40}
\newcommand{\nSizeSmallLat}{12}
\newcommand{\nSizeTtftShareHi}{13}
\newcommand{\nSizeTtftShareLo}{12}
\newcommand{\nSizeTtftShareSmall}{26}

\newcommand{\nSwitchSlowdownMax}{3.2}
\newcommand{\nSwitchSlowdownMin}{1.5}

\newcommand{\nTopCostFullMax}{46.1}
\newcommand{\nTopCostFullMin}{22.5}

\newcommand{\nTopLatFullMax}{17.7}
\newcommand{\nTopLatFullMin}{3.7}

\newcommand{\nTopSwFullMax}{21.4}
\newcommand{\nTopSwFullMin}{8.5}

\newcommand{\nTputRoutedFull}{-2.4}

\newcommand{\nTputTotalFull}{3.8}

\newcommand{\nXrHitNone}{80.4}

\newcommand{\nXrHitSwitch}{52}

\newcommand{\nXrPromptMean}{19}
\newcommand{\nXrPromptPnn}{113}

\newcommand{\nXrSwitchRate}{19}
\newcommand{\nXrSwitchSlowdown}{3.2}
\newcommand{\nXrSwitchesPerSess}{2.3}

\hypersetup{pdfsubject={Preprint}}
\mlsystitlerunning{\textsc{ShadowPrefill}: Keeping Your Next Model Warm for Per-Step Model Routing in Agentic LLM Serving}

\begin{document}

\twocolumn[
\mlsystitle{\sys{}: Keeping Your Next Model Warm for\\ Per-Step Model Routing in Agentic LLM Serving}

\mlsyssetsymbol{equal}{*}
\begin{mlsysauthorlist}
\mlsysauthor{Haoran Qiu}{equal,azure-research}
\mlsysauthor{Rohan Mahapatra}{equal,azure-research}
\mlsysauthor{Rodrigo Fonseca}{azure-research}
\mlsysauthor{Mark Russinovich}{azure}
\end{mlsysauthorlist}

\mlsysaffiliation{azure-research}{Microsoft Azure Research}
\mlsysaffiliation{azure}{Microsoft Azure}

\mlsyskeywords{LLM serving, model routing, KV cache, agents, scheduling}

\vskip 0.1in

\begin{abstract}
Per-step model routers send each LLM call of an agent to the cost-optimal model dynamically within each session.
However, every model switch discards prefix-cache reuse: KV caches are per model, so the new model must prefill the relevant context before it emits a token.
Across representative agentic workloads, including a production coding-agent trace with a mean context of \nGhPromptMean{}K tokens, the prefix hit rate falls from \nHitSameMin--\nHitSameMax\% on same-model calls to \nHitSwitchMin--\nHitSwitchMax\% after a switch, and a switch call takes \nSwitchSlowdownMin--\nSwitchSlowdownMax$\times$ as long.
We present \sys{}, which prefills the session's context on the model the router has not chosen, in capacity that instances serving normal traffic leave idle, so the next switch finds its prefix cached.
\sys{} harvests the per-iteration SLO slack of requests in flight, protects routed calls, spills shadow KV to host memory, and falls back to a cold switch whenever the shadow is incomplete.
With LLMRouter routing 24 model pairs over five agentic benchmark suites, \sys{} cuts switch-call latency by \nTopSwFullMin--\nTopSwFullMax\%, LLM-call latency by \nTopLatFullMin--\nTopLatFullMax\%, and routed token cost by \nTopCostFullMin--\nTopCostFullMax\%, while normal traffic keeps its SLO and energy per request rises by \nEnergyFull\%.
\end{abstract}

]

\begingroup
\renewcommand{\thefootnote}{}
\footnotetext{\textsuperscript{*}Equal contribution.
\textsuperscript{1}Microsoft Azure Research.
\textsuperscript{2}Microsoft Azure.}
\endgroup

\section{Introduction}
\label{sec:intro}

\begin{figure*}[t]
  \centering
  \includegraphics[width=\textwidth]{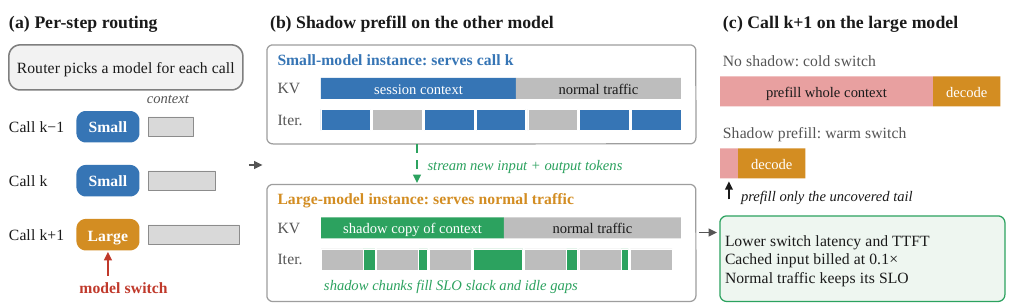}
  \vspace{-6mm}
  \caption{\sys{} overview. (a) A per-step router picks a model for every LLM call of an agent session. (b) While the small model serves call $k$, the large model's instance, which is busy with normal traffic, prefills a shadow copy of the session context in its SLO slack and idle gaps, and receives the new input and output tokens as they appear. (c) When the router switches at call $k{+}1$, the large model finds the prefix cached and prefills only the uncovered tail from the small model.}
  \label{fig:overview}
  \vspace{-3mm}
\end{figure*}

\myparagraph{Motivation}
Agents solve a task through a sequence of LLM calls: plan, call a tool, read the result, edit, test, and repeat.
The calls differ widely in difficulty, so a single model is either too weak for the hard steps or too expensive for the easy ones.
Routers address this mismatch by selecting a model for each call~\cite{ong2024routellm,chen2023frugalgpt,ding2024hybridllm}, and recent systems make the decision per step inside an agent loop~\cite{zhang2025routerr1,dialrouter2026,mtrouter2026,agentrouter2026}.
Commercial products ship the same pattern: GPT-5 routes between a fast and a reasoning model per request~\cite{openai2025gpt5}, and coding agents delegate subtasks to smaller models~\cite{anthropic2025claudecode}.
Across representative agentic workloads, an LLMRouter~\cite{llmrouter2025} policy that routes between two models is \nAccGainLargeMin--\nAccGainLargeMax{}\% more accurate than always using the larger one using less tokens (\Cref{sec:eval:acc}).

\myparagraph{The model switch tax}
Agent calls are long and grow: each call resends the session's history plus new tool output.
Serving engines make this cheap through prefix caching~\cite{kwon2023vllm,gao2024cachedattention,qin2024mooncake}: a call whose prompt extends the previous one prefills only the new tokens, and providers bill cached input at about a tenth of the price.
KV caches, however, belong to one model.
When the router switches, the target model holds no copy of the recent context, or a copy that stops at the last call it served, and must prefill the gap before it emits a token~\cite{liu2026copilot}.
Across representative benchmarks the token-weighted prefix hit rate is \nHitSameMin--\nHitSameMax\% on same-model calls but \nHitSwitchMin--\nHitSwitchMax\% on switch calls, and a switch call takes \nSwitchSlowdownMin--\nSwitchSlowdownMax$\times$ as long as a same-model call (\Cref{sec:bg}).
The switch tax lands on latency, cost, and the flow of the agent loop, and grows with the long, tool-heavy sessions that routing is meant to make affordable.

\myparagraph{Idle LLM serving capacity}
The model the router will switch to is usually already deployed and serving requests from normal traffic.
Such deployments are provisioned for their peak load to avoid SLO violations.
In the Azure production LLM inference traces~\cite{stojkovic2025dynamollm,qiu2025modserve}, reserved GPU pools can only be utilized at {50\% and 39\%} when provisioning for the P95 demand of text and multimodal workloads; even at peak, when whole-instance idle time vanishes, decode iterations are memory-bound and run well within their per-token SLO, leaving \nHeadConvPeak--\nHeadBurstPeak\% of each iteration's latency budget unused (\Cref{sec:bg:idle}).

\myparagraph{Our Work}
We propose \textbf{\sys{}}, which proactively warms up the prefix cache on the model that the router did not choose by harvesting idle capacity (\Cref{fig:overview}).
While the chosen model serves a call, \sys{} prefills the session's context on the other model's instance as low-priority work, and streams the call's input and output tokens to it as they appear.
Upon model switches, the switch call finds most of its prompt cached; otherwise, shadow KV is evicted first and the normal path is unchanged.
Additionally, \sys{} must address these challenges:
\begin{itemize}[leftmargin=*,itemsep=0pt,topsep=1pt]
\item \textbf{[C1] No free GPUs at peak.} Whole instances are rarely idle when load is high (\Cref{sec:bg:idle}), which is when switches are most expensive.
\item \textbf{[C2] Shadow work slows the calls it is meant to speed up.} Shadow chunks share iterations with routed calls and with normal traffic on existing model instances.
\item \textbf{[C3] Long contexts exceed the KV budget.} A 100K-token coding context occupies a large share of a small model's KV pool and is evicted before it is used.
\item \textbf{[C4] Most shadow work can be wasted.} Switches occur on \nGhSwitchRate--\nRrSwitchRate\% of calls (\Cref{sec:bg:routing}), and the router decides only when the call arrives (except for \textit{Plan} mode).
\item \textbf{[C5] No SLO regression.} A switch must never wait for shadow, and normal traffic must keep its SLO.
\end{itemize}
\sys{} answers these with an SLO-slack harvest controller that fills each iteration up to the latency slack of the requests in flight (C1), routed-step protection that limits stretch on iterations carrying routed calls (C2), a capped shadow KV class with a host-memory spill tier (C3), incremental, streamed, and sticky shadow jobs ordered by switch likelihood (C4), and lowest-priority, iteration-granular shadow work with a cold fallback (C5).
It also extends to prefill/decode (PD)-disaggregated clusters.

\myparagraph{Results}
We implement \sys{} on vLLM and evaluate it with LLMRouter~\cite{llmrouter2025} as the per-step router on 24 model pairs over five agentic benchmarks, including SWE-bench~\cite{jimenez2024swebench} and the GitHub Copilot coding-agent trace~\cite{liu2026copilot}, with production LLM inference traffic from Azure~\cite{stojkovic2025dynamollm}.
\sys{} reduces switch-call latency by \nBenchSwFull\% on the benchmarks and \nGhcpSwFull\% on coding sessions, switch-call TTFT by \nBenchTtftFull\% and \nGhcpTtftFull\%, mean LLM-call latency by \nBenchLatFull\% and \nGhcpLatFull\%, and routed token cost by \nBenchCostFull\% and \nGhcpCostFull\%.
It captures \nBenchLatCapture\% of the latency gain of an oracle that makes switches free, without extra GPUs, while energy per request rises by \nEnergyFull\% and normal traffic keeps its SLO.

\myparagraph{Contributions}
In summary, our main contributions are:
\begin{itemize}[leftmargin=*,nosep,topsep=0pt]
\vspace{-0.4\baselineskip}
\item A characterization of per-step routing in agent workloads that quantifies the model switch tax (\Cref{sec:bg}).
\item \sys{}, the first system that warms a routed-to model's prefix cache by harvesting the idle capacity of instances serving normal traffic, with mechanisms for each challenge and a PD-disaggregated variant (\Cref{sec:design}).
\item A comprehensive evaluation over 24 model pairs across five agentic benchmarks and traces that demonstrate the benefits from \sys{} (\Cref{sec:eval,sec:testbed}).
\end{itemize}

\section{Background and Characterization}
\label{sec:bg}

\begin{figure*}[t]
  \centering
  \includegraphics[width=\textwidth]{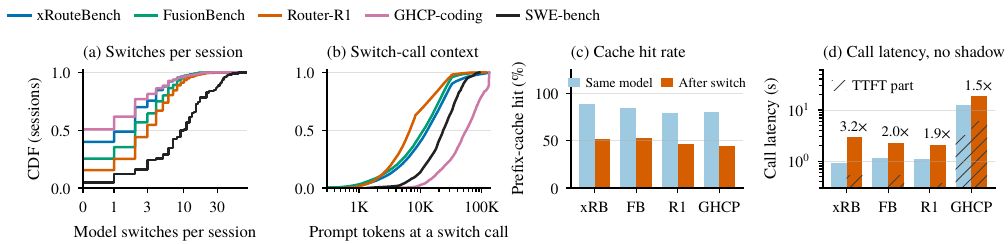}
  \vspace{-6mm}
  \caption{Per-step routing and its switch tax. (a) Model switches per routed session. (b) Prompt length of switch calls. (c) Token-weighted prefix-cache hit rate of same-model and switch calls. (d) Mean call latency (hatched: time to first token); labels are switch-call slowdown.}
  \label{fig:char}
  \vspace{-3mm}
\end{figure*}

\subsection{Per-step Routing in Agent Loops}
\label{sec:bg:routing}

\myparagraph{Model routers}
A per-step router observes each LLM call of a session and selects one of several deployed models for it.
We adopt LLMRouter~\cite{llmrouter2025} as the router.
Its benchmarks evaluate every query on every candidate model, so a routing policy can be replayed exactly; we use its accuracy-per-cost policy, which sends each step to the model with the highest score per dollar, and restrict the candidates to a two-model pair, a small model \Smodel and a large model \Lmodel.
Two models is the common deployment of a cheap and a capable tier~\cite{openai2025gpt5,anthropic2025claudecode}.

\myparagraph{Workloads}
We study five routed workloads, as summarized in \Cref{tab:traces}:
xRouteBench~\cite{xroutebench}, FusionBench~\cite{feng2025fusionbench}, and Router-R1~\cite{zhang2025routerr1} are routing benchmarks; we group consecutive queries of a task into sessions and grow the context as an agent loop does, appending each step's input, output, and a tool result (lognormal, median 1K tokens) to the history.
GHCP-coding takes its token sizes from one day of the public GitHub Copilot coding-agent trace~\cite{ghcpdataset2026}: each session is one user turn of up to 60 agent calls, and each call carries its recorded prompt, which already includes the history and tool outputs.
GHCP itself rarely changes model, so its routing decisions come from routing FusionBench's coding tasks (HumanEval, MBPP) within the same pair.
SWE-bench~\cite{jimenez2024swebench} sessions are SWE-agent trajectories (57 calls on average) whose calls LLMRouter's matrix-factorization router routes between SWE-agent-LM-7B and SWE-agent-LM-32B or Qwen2.5-72B; we replay them on the GPU testbed (\Cref{sec:testbed}).
Per-step routing is frequent: \nXrSwitchRate--\nRrSwitchRate\% of calls switch model and a session switches \nGhSwitchesPerSess--\nRrSwitchesPerSess{} times on average (\Cref{fig:char}a); SWE-bench sessions switch 13 times.
Coding contexts are long: the mean switch call carries \nGhPromptMean{}K tokens on GHCP and 27K on SWE-bench, 3--6$\times$ and 1.3--2.6$\times$ the routing benchmarks (\Cref{fig:char}b).
The router is \nAccGainLargeMin--\nAccGainLargeMax{} points more accurate than always using the pair's more expensive model, because the two models are right on different queries, yet it saves only \nNoneVsLargeMin--\nNoneVsLargeMax\% of that model's token cost, partly because every switch forfeits the cached-input discount (\Cref{sec:eval:acc}).

\begin{table}[t]
  \centering
  \vspace{-10pt}
  \caption{Routed workloads. Prompt sizes in tokens; hit rates are token-weighted over all calls and over switch calls.}
  \label{tab:traces}
  \small
  \setlength{\tabcolsep}{2.4pt}
  \begin{tabular}{@{}lrrrrrr@{}}
    \toprule
    Trace & Pairs & Switch & Sw./sess. & Prompt & p99 & Hit (sw.) \\
    \midrule
    xRouteBench & 5 & \nXrSwitchRate\% & \nXrSwitchesPerSess & \nXrPromptMean{}K & \nXrPromptPnn{}K & \nXrHitNone{} (\nXrHitSwitch) \\
    FusionBench & 5 & \nFbSwitchRate\% & \nFbSwitchesPerSess & \nFbPromptMean{}K & \nFbPromptPnn{}K & \nFbHitNone{} (\nFbHitSwitch) \\
    Router-R1 & 6 & \nRrSwitchRate\% & \nRrSwitchesPerSess & \nRrPromptMean{}K & \nRrPromptPnn{}K & \nRrHitNone{} (\nRrHitSwitch) \\
    GHCP-coding & 6 & \nGhSwitchRate\% & \nGhSwitchesPerSess & \nGhPromptMean{}K & \nGhPromptPnn{}K & \nGhHitNone{} (\nGhHitSwitch) \\
    SWE-bench & 6 & 26\% & 10.9 & 36K & 93K & 95.2 (65) \\
    \bottomrule
  \end{tabular}
  \vspace{-3mm}
\end{table}

\subsection{The Model Switch Tax}
\label{sec:bg:tax}

\myparagraph{Prefix caching is per model}
Serving engines keep the KV of recent prompts and reuse it when a new prompt shares a prefix~\cite{kwon2023vllm,gao2024cachedattention,qin2024mooncake}.
In an agent loop each call extends the previous one, so a session that stays on one model prefills only the new tool output and input: same-model calls hit \nHitSameMin--\nHitSameMax\% of their prompt tokens (\Cref{fig:char}c).
KV is a function of the model's weights, so the cache cannot be shared across models.
A switch call finds either nothing on the target or a copy that ends where the session last left that model, and its hit rate falls to \nHitSwitchMin--\nHitSwitchMax\%.
Same observation is made in a study on Github Copilot model switch behaviors~\cite{liu2026copilot}.

\myparagraph{The tax is paid in latency, cost, and flow}
The lost prefix must be prefilled before the first token.
A switch call takes \nXrSwitchSlowdown$\times$ as long as a same-model call on xRouteBench and \nGhSwitchSlowdown$\times$ on GHCP-coding (\Cref{fig:char}d); time to first token (TTFT), the stall a user or agent loop waits through before anything streams, accounts for most of the difference.
Providers bill cached input at about a tenth of the uncached price, so every lost hit also raises the token consumption for the end user, and the extra prefill occupies GPUs that serve other users.
Switch-aware routers~\cite{dialrouter2026,mtrouter2026} avoid switches and give up accuracy; \sys{} lowers their price.

\begin{figure}[t]
  \centering
  \includegraphics[width=\columnwidth]{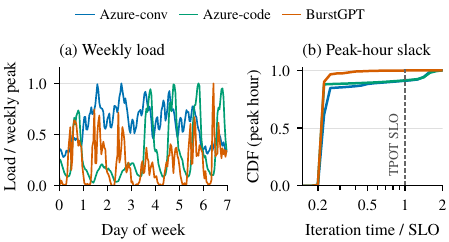}
  \vspace{-8mm}
  \caption{Idle capacity of LLM deployments. (a) Load over one week, relative to the trace's peak. (b) Iteration time relative to the TPOT SLO at the peak hour.
  Results are for Qwen2.5-7B (one H100 GPU per instance) and Llama-3.3-70B (eight H100 GPUs per instance), with four instances per pool and target utilization $\rho=0.7$.}
  \label{fig:idle}
  \vspace{-4mm}
\end{figure}

\subsection{Idle Capacity in Production Deployments}
\label{sec:bg:idle}

\myparagraph{Deployments are provisioned for peak}
The model a session switches to is already deployed and busy serving other users' requests, which we call normal traffic.
Clusters are typically sized for the peak hour to reserve buffer for bursts and avoid SLO violations, so off-peak hours leave capacity unused.
We replay one week of the Azure LLM inference traces for conversation and code~\cite{stojkovic2025dynamollm} and BurstGPT~\cite{wang2024burstgpt} on a pool provisioned for its peak (\Cref{fig:idle}a).
Load swings by 3--10$\times$ within a day.
Instances serving code traffic are completely idle for \nIdleCodeMed\% of the median hour and \nIdleCodeTrough\% of the trough hour, but only \nIdleCodePeak\% of the peak hour, and conversation traffic, which is steadier, leaves no whole-instance idle time in any window.

\myparagraph{Iterations have SLO slack even at peak}
Whole-instance idleness disappears exactly when switches are slowest.
A finer-grained resource remains: a continuous-batching iteration is dominated by memory-bound decode and finishes well within the time-per-output-token (TPOT) SLO of the requests it serves.
At the peak hour, the median iteration uses about a quarter of the TPOT SLO, and \nHeadConvPeak--\nHeadBurstPeak\% of the per-iteration latency budget is unused (\Cref{fig:idle}b).
A prefill chunk added to such an iteration delays every decode in it~\cite{agrawal2024taming}, but as long as the iteration stays within the slack of each request in flight, no request misses its SLO.
This slack is the capacity \sys{} harvests.

\section{Design}
\label{sec:design}

\sys{} sits between the router and the serving engines of a model pair.
For every routed call it creates or extends a \emph{shadow job} on the model the router did not choose, and each engine runs shadow jobs as a lowest-priority work class.
This section states the goal and the mechanisms to address the challenges: Shadow jobs (\textbf{[C4]}, \Cref{sec:design:jobs}), harvesting (\textbf{[C1]}, \textbf{[C2]}, \Cref{sec:design:harvest}), Shadow KV management (\textbf{[C3]}, \Cref{sec:design:kv}), and safety (\textbf{[C5]}, \Cref{sec:design:safety}).

\subsection{Goal and Shadow Jobs}
\label{sec:design:jobs}

\myparagraph{Goal}
When a session switches from model $A$ to model $B$ at call $k{+}1$, the prompt of that call consists of the session's history $h_k$ (all prompts and outputs through call $k$) plus the new input.
\sys{} aims to have $h_k$ resident in $B$'s prefix cache when call $k{+}1$ arrives, so that the switch call prefills only the new input, as a same-model call does.
It may spend only capacity that would otherwise go unused (by harvesting idle capacity), and it must never delay a switch call or cause normal traffic to miss its SLO.

\myparagraph{Incremental shadow jobs}
A shadow job for session $s$ on model $B$ records the prefix of $h$ that $B$ already holds and the tokens still missing.
When call $k$ arrives at $A$, the manager appends call $k$'s new input to the job immediately, and appends $A$'s output tokens once $A$ generates them (output streaming).
$B$ therefore never re-prefills what it already has: after a switch back from $B$ to $A$ and forward again, the job covers only the delta.
The shadow always trails the routed model by at least the output of the current call; the switch call prefills that tail itself as it would do at each switch.

\myparagraph{Sticky model placement}
Each agentic session keeps a sticky instance per model~\cite{agarwal2026symphony}, and the shadow job runs on the sticky instance of the other model.
The router sends each call to the instance of the chosen model with the lowest estimated prefill completion time, counting only uncached tokens.
A switch call therefore lands on the instance that holds its shadow, unless that instance is so busy that a cold prefill elsewhere would finish first.

\myparagraph{Ordering by switch likelihood (C4)}
The engine runs pending shadow jobs in order of their predicted switch probability, estimated per task and current model from past sessions, so the capacity goes first to the sessions most likely to switch.
Gating jobs below a probability threshold reduces waste further but loses switches the predictor misses; we evaluate it as the \predictive variant in \Cref{sec:eval}.

\subsection{Harvesting Capacity}
\label{sec:design:harvest}

\myparagraph{Where shadow work runs}
Each iteration of a continuous-batching engine schedules, in priority order, all running decodes, in-flight prefill chunks, and admissions from the waiting queue, under a token budget.
\sys{} adds shadow chunks last, and only if the waiting queue is empty and no admission was blocked by memory.

\myparagraph{Idle instances}
When an instance has no other work, it runs shadow-only iterations.
Each such iteration is capped at $\beta$ times the TPOT SLO ($\beta{=}0.5$), which bounds how long a newly arriving request waits behind it.

\myparagraph{SLO-slack controller (C1)}
At peak, instances are rarely idle, so \sys{} piggybacks shadow chunks on normal iterations, sized by the latency slack of the requests in flight.
The controller uses only what a production scheduler observes: arrival times, first-token times, and iteration counts.
A decoding request $r$ that received its first token at $f_r$ and has completed $k_r$ iterations meets a TPOT SLO $T_r$ with margin $m$ if $t - f_r \le (1{-}m)\,T_r\,k_r$, so the next iteration may last up to
\begin{equation}
  \sigma_r = (1{-}m)\,T_r\,(k_r{+}1) - (t - f_r).
\end{equation}
A prefilling request has slack $(1{-}m)F_r - (t - a_r) - \hat{p}_r$, where $F_r$ is its TTFT SLO, $a_r$ its arrival, and $\hat{p}_r$ the predicted time of its remaining chunks.
The iteration budget is the smallest slack in the batch, capped at $\beta T$; the controller then adds shadow tokens while the latency model predicts the iteration to stay within budget (\Cref{alg:slack}).
Requests that are ahead of their SLO, which at the median iteration is most of them (\Cref{fig:idle}b), lend their slack to shadow work; requests close to their deadline block it.

\myparagraph{Routed-step protection (C2)}
Routed calls have no SLO slack to lend: any stretch of an iteration that carries one adds directly to the agent's latency, including that of switch calls the shadow is meant to speed up.
When a routed call is prefill/decoding on the instance, the controller instead spends a small credit that accrues at $\delta$ times the duration of memory-bound, decode-only iterations ($\delta{=}0.025$), capped by the budget.
The long-run slowdown of routed calls is therefore at most $\delta$, while iterations that carry only normal traffic harvest their full slack.
Without the SLO controller, \harvest uses the same credit on every busy iteration ($\delta{=}0.1$).

Sensitivity studies on parameters are in \Cref{sec:eval:where,app:sensitivity}.

\begin{algorithm}[t]
  \caption{Shadow tokens added to one iteration.}
  \label{alg:slack}
  \small
  \begin{algorithmic}[1]
    \REQUIRE batch $\mathcal{B}$ of decodes and prefill chunks; shadow queue $Q$
    \IF{waiting queue nonempty \OR admission blocked}
      \STATE \textbf{return} 0
    \ELSIF{$\mathcal{B} = \emptyset$}
      \STATE $b \gets \beta T$ \hfill{\footnotesize idle instance}
    \ELSIF{$\mathcal{B}$ holds a routed call}
      \STATE $b \gets \min(\textit{credit}, \beta T)$ \hfill{\footnotesize C2}
    \ELSE
      \STATE $b \gets \min(\beta T, \min_{r \in \mathcal{B}} \sigma_r)$ \hfill{\footnotesize C1}
    \ENDIF
    \STATE $n \gets \max\{n : \hat{t}(\mathcal{B} + n\ \text{shadow tokens}) \le b\}$
    \STATE $n \gets \min(n,\ \text{free shadow KV})$
    \IF{$n < n_{\min}$}
      \STATE \textbf{return} 0
    \ENDIF
    \STATE take $n$ tokens from jobs in $Q$ by switch probability
    \STATE \textbf{return} $n$
  \end{algorithmic}
\end{algorithm}

\subsection{Shadow KV Management}
\label{sec:design:kv}

\myparagraph{A capped, evict-first class (C3)}
Shadow KV is useful only if it survives until the switch, but it must not displace the cache entries that make same-model calls fast.
Shadow entries form their own cache class, limited to 25\% of the KV pool.
A shadow chunk may use free blocks or blocks of cache entries idle for more than 5\,s.
Eviction removes shadow entries before normal ones and is LRU within each class.
Configuration sensitivity studies are in \Cref{sec:eval:where}.

\myparagraph{Host-memory spill tier}
\sys{} spills evicted session entries, shadow and normal, to host DRAM and reloads them over PCIe when a call hits them, as KV offloading systems do for a single model~\cite{gao2024cachedattention,cheng2025lmcache}.
Reloading a prefix is far cheaper than recomputing it, so the tier keeps shadow work from being lost to HBM pressure; it also helps same-model calls whose prefixes were evicted, which we evaluate separately as the \hosttier variant (\Cref{sec:eval}).
Host tier is needed especially for long-context agents (e.g., coding~\cite{liu2026copilot}): Gemma-7B uses full multi-head attention and stores 458\,KB of KV per token, so a single 100K-token agent context fills most of its H100's KV pool, and its shadow is evicted long before the switch.

\subsection{Safety and Fallback}
\label{sec:design:safety}

\myparagraph{No regression by construction (C5)}
Shadow work never blocks a routed call or a normal request: it runs only when nothing is waiting, it is bounded per iteration by the budget above, and it is preempted at iteration granularity, so an arriving request waits at most one capped iteration.
A switch call never waits for a shadow.
If its shadow is partial, it uses the cached prefix and prefills the uncovered tail; if the shadow is absent or was evicted, it runs exactly as without \sys{}.
The price of a wrong prediction is the harvested capacity it consumed, plus the energy of that computation (\Cref{sec:eval:overheads}).

\subsection{Prefill/Decode Disaggregation Extension}
\label{sec:design:pd}

PD disaggregation runs prefill and decode on separate instances and moves the KV between them~\cite{patel2024splitwise,qin2024mooncake}.
A switch there costs both a cold prefill on the target's prefill (P) instances and a larger KV transfer to its decode (D) instance.
\sys{} runs shadow jobs on the session's sticky P instance of the other model when its queue is empty and background requests have TTFT slack, in iterations of at most 0.1\,s.
P instances are provisioned for prefill-heavy peaks and idle often, so shadow work finds more capacity than in a colocated cluster.
The sticky D instance of the other model then pulls the shadow KV in 4K-token pushes whenever its inbound link carries no normal transfer, into free decode memory that is reclaimed first.
At the switch, the P instance prefills only the tail and the D instance receives only the tokens it lacks.

\section{Implementation}
\label{sec:impl}

\myparagraph{Router and sessions}
We build routed sessions from LLMRouter's benchmark matrices~\cite{llmrouter2025} by replaying the accuracy-per-cost policy over each model pair (\Cref{tab:pairs}).
A session builder grows each step's prompt from the session history, the dataset input, and optionally sampled tool output; it compacts the history into a 1K-token summary when the context reaches the model's window.

\myparagraph{Serving simulator}
We implement on top of Vidur~\cite{agrawal2024vidur}, a high-fidelity LLM inference simulator, with additionally 2,700 lines of Python.
The resulting simulator is a discrete-event model of a cluster of continuous-batching instances.
Each iteration’s latency is grounded by a per-layer profiled model of kernels (GEMMs, prefill attention, decode attention, and tensor-parallel all-reduces) fitted to each model's H100 operator profiles.
Model instances run chunked prefill with a default 2,048-token budget and up to 256 sequences, reserve KV cache at admission, and maintain a prefix cache of session contexts and finished requests.

\begin{figure*}[t]
  \centering
  \includegraphics[scale=0.92]{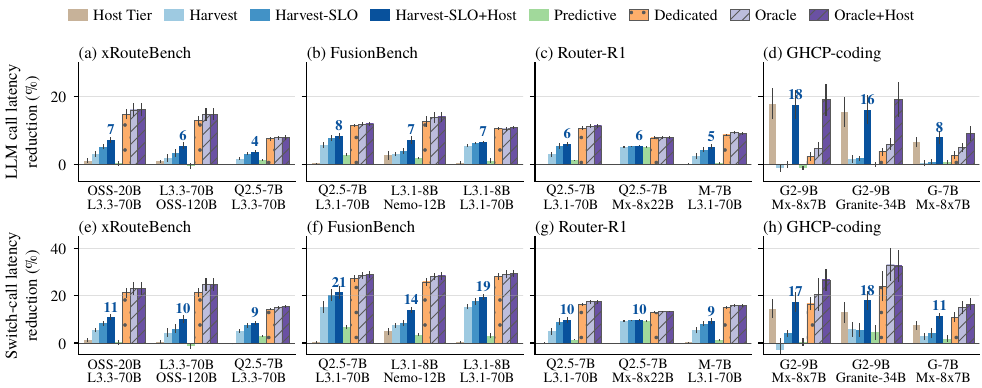}
  \vspace{-2mm}
  \caption{Reduction in time-to-last-token latency of (a--d) all LLM calls and (e--h) switch calls relative to \nosh for each trace. Labels indicate the improvements from \full. Error bars: $\pm$1 SE over 3 seeds $\times$ 3 windows.}
  \label{fig:lat}
  \vspace{-2mm}
\end{figure*}

\myparagraph{vLLM prototype}
We build a prototype to validate the core mechanism of \sys{} in a real serving system.
For each model pair, we run two vLLM~\cite{kwon2023vllm} instances, one per model, with prefix caching and priority scheduling enabled.
A router selects the model for each call in an agentic session, and a router-side manager implements a request-level version of \harvest.
After a call completes on one instance, the manager submits its prompt and response to the other instance as low-priority shadow work, allowing that model to compute and retain its own KV for reuse after a subsequent model switch.
For learned model routing, we follow LLMRouter's matrix-factorization methodology~\cite{llmrouter2025} and retrain a router for each model pair.
The implementation is approximately 840 lines of Python code and require no modifications to vLLM source code.
Further details are in \Cref{app:testbed-setup}.

\section{Evaluation}
\label{sec:eval}

We evaluate \sys{} through trace-driven simulation in this section and GPU testbed experiments in \Cref{sec:testbed}.

\subsection{Setup}
\label{sec:eval:setup}

\myparagraph{Model pairs and benchmarks}
We simulate 22 model pairs: 5 on xRouteBench, 5 on FusionBench, 6 on Router-R1, and 6 on GHCP-coding (\Cref{tab:traces}; full list in \Cref{tab:pairs}); the GPU testbed adds two SWE-bench pairs (\Cref{sec:testbed}). The simulated pairs span dense and MoE models from 7B to 141B parameters, same- and cross-family pairs, and the Qwen, Llama, Mistral, Gemma, GPT-OSS, and Granite families.
The appendix shows all model pairs (\Cref{tab:pairs} and \Cref{tab:testbed-labels}).

\myparagraph{Production traffic}
Each model runs on 4 instances, each using the smallest tensor-parallel degree among 1, 2, 4, and 8 H100 GPUs at which weights take at most half the HBM.
Each instance serves normal traffic replayed from the Azure LLM inference conversation trace~\cite{stojkovic2025dynamollm} by default.
Pools are provisioned for the trace's peak hour: the normal-traffic rate is $\rho{=}0.7$ of the capacity at which 90\% of requests meet both SLOs at peak, and we run three one-hour windows (peak, median, and trough) at their relative trace load.
The SLOs are TTFT $\le \max(1\,\text{s}, 5\times$ isolated prefill time$)$ and TPOT $\le \max(25\,\text{ms}, 5\times$ batch-1 decode time$)$.

\myparagraph{Measurement}
Each run measures 600\,s (1,800\,s for GHCP-coding) after a 120\,s warmup and drains until every measured session finishes.
We run 3 seeds $\times$ 3 windows per configuration and compare each approach with \nosh on the same seed, window, and pair, over the calls that complete under every approach; error bars are $\pm$1 standard error over these runs.
Token cost uses each benchmark's per-model prices, with cached input billed at 10\% of uncached.

\begin{table}[t]
  \centering
  \vspace{-10pt}
  \caption{Approaches compared.}
  \label{tab:approaches}
  \small
  \setlength{\tabcolsep}{3pt}
  \begin{tabularx}{\columnwidth}{@{}lX@{}}
    \toprule
    Approach & Behavior \\
    \midrule
    \nosh & Cold switch; the baseline. \\
    \hosttier & \nosh with the host-DRAM KV tier. \\
    \harvest & Shadow jobs under a fixed credit ($\delta{=}0.1$). \\
    \harvestslo & Shadow jobs under the SLO-slack controller. \\
    \full & \harvestslo with the host tier. \\
    \predictive & \harvest, gated on switch probability $\ge 0.25$. \\
    \dedicated & One extra instance per model runs all shadow work (+\nDedicatedGpus\% GPUs). \\
    \eager & Shadow work as normal-priority requests. \\
    \oracle(+\textcolor{darkgreen}{Host}) & Switch call's prefix cached for free. \\
    \bottomrule
  \end{tabularx}
  \vspace{-3mm}
\end{table}

\begin{figure*}[t!]
  \centering
  \includegraphics[scale=0.92]{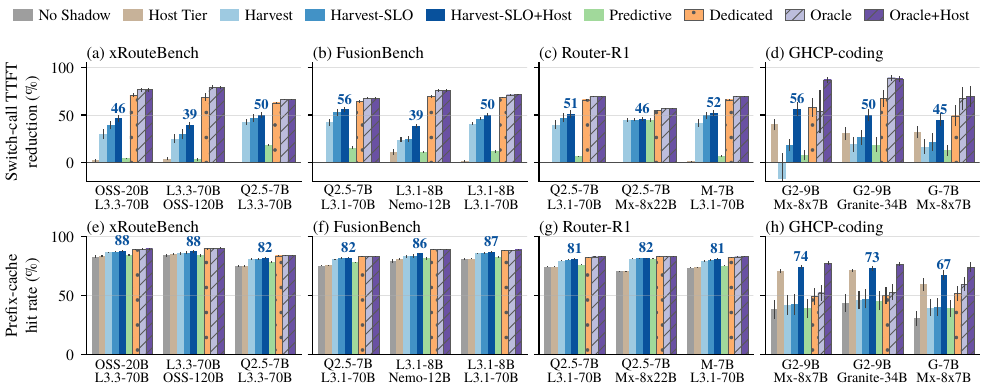}
  \vspace{-2mm}
  \caption{(a--d) Reduction in switch-call TTFT relative to \nosh. (e--h) Token-weighted prefix-cache hit rate of all LLM calls. Labels indicate the percentage improvement or cache hit performance from \full.}
  \label{fig:ttft}
  \vspace{-3mm}
\end{figure*}

\myparagraph{Baselines, ablations, and oracle}
\Cref{tab:approaches} lists the different system configurations.
\hosttier isolates what the host tier alone does for same-model calls; \harvest and \harvestslo isolate the two controllers; \full is the complete design.
\dedicated spends extra GPUs instead of harvesting, and \eager uses the engine's normal request path, as prior cache-warming systems do for a single model~\cite{cachescout2026}.
\oracle and \oraclehost bound what any warming scheme can achieve without and with the host tier.
Figures omit \eager, which significantly slows LLM calls.

\begin{figure*}[h]
  \centering
  \includegraphics[scale=0.9]{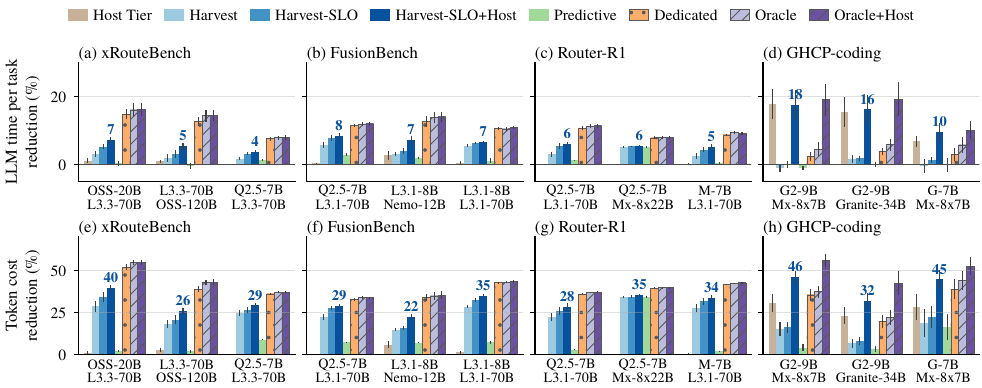}
  \vspace{-2mm}
  \caption{(a--d) Reduction in LLM time per task (sum of a session's call latencies) and (e--h) reduction in routed token cost, relative to \nosh. Labels indicate the percentage improvement from \full.}
  \label{fig:taskcost}
  \vspace{-3mm}
\end{figure*}

\subsection{Latency and Cost Comparison}
\label{sec:eval:e2e}

\myparagraph{Switch calls}
\sys{} improves the switch tax directly (\Cref{fig:lat}e--h).
\full cuts switch-call latency by \nBenchSwFull\% on average over the 16 benchmark pairs and by \nGhcpSwFull\% over the 6 GHCP-coding pairs, and by up to \nAllSwFullMax\% on a single pair.
\oracle bounds the achievable reduction at \nBenchSwOracle\% and \nGhcpSwOracleHost\% (\oraclehost on GHCP), and \full reaches \nBenchSwCapture\% and \nGhcpSwCapture\% of it without extra GPUs.
On GHCP-coding a cold switch call takes \nGhcpSwAbsNone\,s, and \sys{} saves \nGhcpSwSaveFull\,s of it.
The improvement comes mostly from the time to generate the first token: switch-call TTFT falls by \nBenchTtftFull\% and \nGhcpTtftFull\% (\Cref{fig:ttft}a--d).

\myparagraph{All LLM calls}
Switch calls are \nXrSwitchRate--\nRrSwitchRate\% of all LLM calls, so the mean latency of all LLM calls falls by \nBenchLatFull\% on the benchmarks and \nGhcpLatFull\% on GHCP-coding (\Cref{fig:lat}a--d), against \oracle bounds of \nBenchLatOracle\% and \nGhcpLatOracleHost\%.

\myparagraph{Cache hit rate}
\Cref{fig:ttft}e--h shows the benefits of \sys{} on cache reuse.
\full improves the token-weighted hit rate of all calls from \nBenchHitNone\% to \nBenchHitFull\% on the benchmarks and from \nGhcpHitNone\% to \nGhcpHitFull\% on GHCP-coding, close to \oraclehost (\nBenchHitOracleHost\% and \nGhcpHitOracleHost\%).

\begin{figure*}[t!]
  \centering
  \includegraphics[scale=0.9]{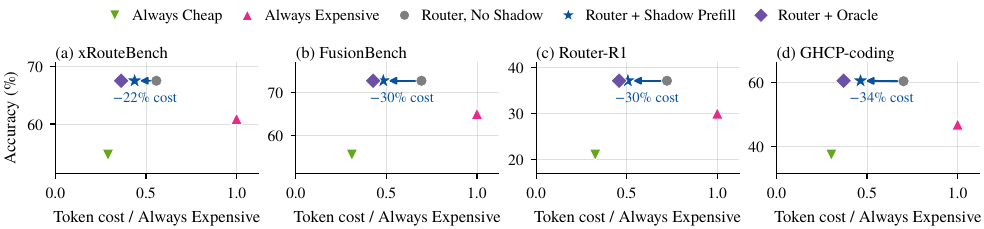}
  \vspace{-4mm}
  \caption{Accuracy versus routed token cost, relative to \alwaysexp, averaged over each trace's model pairs. Arrows mark the cost reduction from \sys{} (\full) at equal accuracy.}
  \label{fig:acccost}
  \vspace{-3mm}
\end{figure*}

\begin{figure*}[t!]
  \centering
  \includegraphics[scale=0.9]{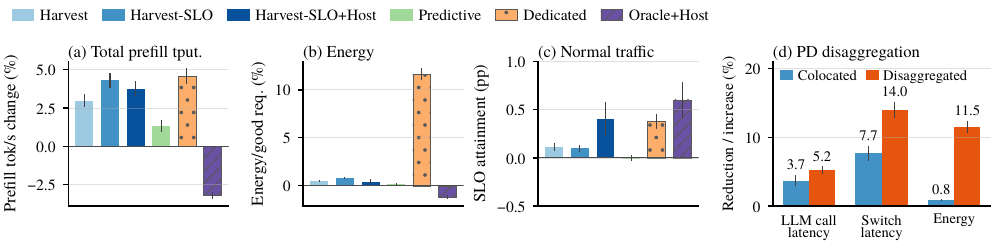}
  \vspace{-4mm}
  \caption{Overheads over all 22 simulated pairs, relative to \nosh. (a) Total prefill throughput, including shadow tokens. (b) Energy per request meeting its SLO. (c) Normal-traffic SLO attainment. (d) \harvestslo on colocated and P/D-disaggregated clusters.}
  \label{fig:overheads}
  \vspace{-3mm}
\end{figure*}

\myparagraph{Ablation on each mechanism}
The fixed-credit \harvest controller reaches \nBenchSwHarvest\% on the benchmarks: it spends the same small credit on every busy iteration, whatever the slack of the requests in it.
The SLO-slack controller lifts this to \nBenchSwSlo\%, and the host tier to \nBenchSwFull\%.
On GHCP-coding, \harvestslo reaches only \nGhcpSwSlo\%, because a 63K-token shadow on a small model's KV pool is evicted before the switch; the host tier keeps it, and \full reaches \nGhcpSwFull\%.
\hosttier alone barely helps the benchmarks (\nBenchLatHostTier\% LLM-call latency), whose contexts fit in HBM.
On GHCP it recovers evicted prefixes of same-model calls too and cuts LLM-call latency by \nGhcpLatHostTier\%, but shadow work on top of it is what fixes switch calls: TTFT falls by \nGhcpTtftHostTier\% with \hosttier and \nGhcpTtftFull\% with \full, and token cost by \nGhcpCostHostTier\% and \nGhcpCostFull\%.
\predictive saves less than \harvest (\nBenchSwPredictive\% \vs \nBenchSwHarvest\%): the router decides per call, and a gate on predicted switch probability skips many switches that do happen.
\dedicated approaches \oracle (\nBenchSwDedicated\%), but needs \nDedicatedGpus\% more GPUs and \nEnergyDedicated\% more energy per request (\Cref{fig:overheads}).
\eager (not shown) issues shadow work as ordinary requests and slows LLM calls by \nBenchLatEagerUp\% on the benchmarks and \nGhcpLatEagerUp\% on GHCP-coding.

\myparagraph{Task completion time}
An agent task issues its calls in sequence, so the LLM time per task falls with the per-call latency (\Cref{fig:taskcost}a--d): by \nBenchTaskFull\% across the agentic benchmarks and by \nGhcpTaskFull\% on GHCP-coding traces, and by up to \nAllTaskFullMax\% on the model pair of Gemma-2-9B and Mixtral-8x7B.

\myparagraph{User-side token cost}
Every token a switch call finds cached is billed at a tenth of the price~\cite{anthropic2025claudecode}, so \sys{} cuts the routed token bill by \nBenchCostFull\% on the benchmarks and \nGhcpCostFull\% on GHCP-coding, and by up to \nAllCostFullMax\% (\Cref{fig:taskcost}e--h) due to its longer prefix.

\subsection{Accuracy and Cost of Routing}
\label{sec:eval:acc}

\sys{} changes the cost of routing without changing its decisions.
\Cref{fig:acccost} places the router, with and without \sys{}, against \alwayscheap and \alwaysexp, which send every call to the pair's cheaper or pricier model by price.
The router is \nAccGainLargeMin--\nAccGainLargeMax{} points more accurate than \alwaysexp, since the two models answer different queries correctly~\cite{llmrouter2025}.
Without \sys{} it saves \nNoneVsLargeMin--\nNoneVsLargeMax\% of \alwaysexp's cost; with \sys{} it saves \nCostVsLargeMin--\nCostVsLargeMax\%, cutting the router's own bill by \nAccCostCutMin--\nAccCostCutMax\% at the same accuracy.
On GHCP-coding, routing with \sys{} is \nGhAccGainLarge{}\% more accurate than \alwaysexp at \nGhCostVsLarge\% lower cost.

\subsection{Overheads and Disaggregation}
\label{sec:eval:overheads}

\myparagraph{Token throughput}
\sys{} moves prefill work from the critical path into harvested idle capacity: under \full, total prefill throughput, including shadow tokens, rises by \nTputTotalFull\% (\Cref{fig:overheads}a), while the prefill throughput of routed and normal requests alone changes by \nTputRoutedFull\%, because switch calls find their prefix cached.
Decode throughput is unchanged.

\myparagraph{Energy and normal traffic}
Shadow work runs on GPUs that are powered anyway, in slack that is mostly memory-bound, so energy per good request rises by only \nEnergyFull\% under \full and \nEnergySlo\% under \harvestslo, compared with \nEnergyDedicated\% for \dedicated (\Cref{fig:overheads}b).
Normal-traffic SLO attainment does not drop under any harvesting approach; it rises by \nBgSloSlo--\nBgSloFull{} points (\Cref{fig:overheads}c), because switch calls that hit the cache leave the instances sooner.

\begin{figure*}[t]
  \centering
  \includegraphics[scale=0.9]{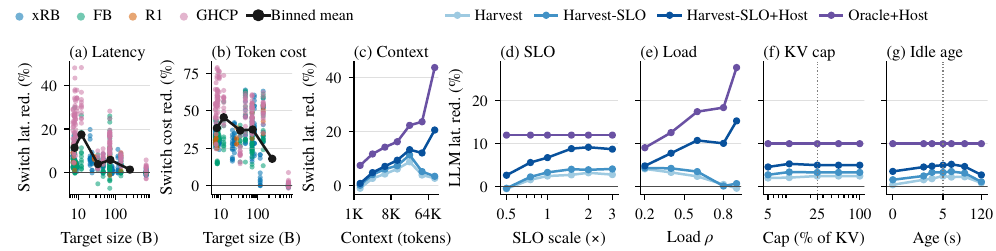}
  \vspace{-4mm}
  \caption{Where the benefit comes from. Switch-call (a) latency and (b) cost reduction of \full versus the size of the target model (points: pair and target; line: call-weighted bin mean). (c) Switch-call latency reduction versus context length. LLM-call latency reduction versus (d) SLO scale, (e) load $\rho$, (f) shadow KV cap, and (g) cold-entry age (dotted: default).}
  \label{fig:where}
  \vspace{-3mm}
\end{figure*}

\providecommand{\testbedfigdir}{figures/gpu}
\begin{figure*}[h]
  \centering
  \includegraphics[scale=0.9]{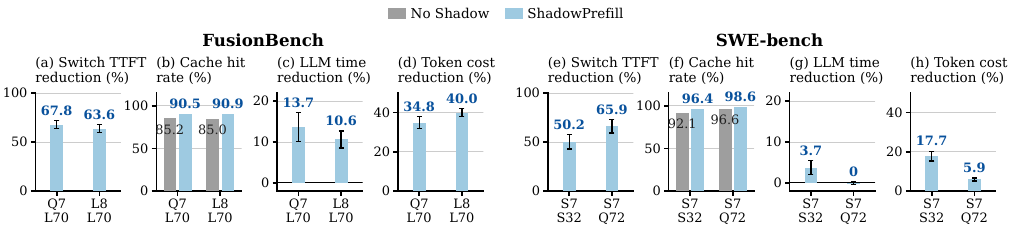}
  \vspace{-5mm}
  \caption{Switch-call TTFT reduction, prefix-cache hit rate, LLM time per task reduction, and token cost reduction relative to \nosh.}
  \label{fig:testbed}
  \vspace{-3mm}
\end{figure*}

\myparagraph{Prefill/Decode disaggregation}
On a disaggregated cluster, each model's 4 instances split into prefill and decode instances, provisioned per trace for the same SLOs.
Prefill instances idle for over half of each run, so \harvestslo finds more capacity: across all pairs it cuts LLM-call latency by \nPdLat\% and switch-call latency by \nPdSw\%, compared with \nColoLat\% and \nColoSw\% colocated (\Cref{fig:overheads}d), and reaches \nPdCapture\% of the disaggregated oracle's gain on the benchmarks.
The price is energy: colocated shadow work rides on iterations that run anyway, whereas disaggregated shadow work wakes otherwise-idle prefill GPUs, raising energy per good request by \nPdEnergy\% (\nColoEnergy\% colocated).

\subsection{Where the Benefit Comes From}
\label{sec:eval:where}

\myparagraph{Target model size}
\Cref{fig:where}a--b breaks the switch-call benefit down by the model a switch goes to, over \nSizePairs{} screened pairs.
Latency falls by \nSizeSmallLat\% for targets under 15B, \nSizeMidLat\% for 15--50B, and \nSizeBigLat\% above 50B.
Shadow Prefill raises the switch-call hit rate by \nSizeHitGainLo--\nSizeHitGainHi{} points at every size up to 141B, but the prefill it removes is \nSizeTtftShareSmall\% of a cold switch call on small targets and \nSizeTtftShareLo--\nSizeTtftShareHi\% on larger ones, whose decode dominates.
DeepSeek-R1 (671B) gains \nSizeHugeHitGain{} points: its long reasoning outputs (\nSizeOutHuge{} tokens on average) leave its instances little slack.
Cost falls by \nSizeSmallCost\%, \nSizeMidCost\%, and \nSizeBigCost\% in the same bins, as output tokens take a growing share of the bill.

\myparagraph{Context length}
The longer the context before a switch, the more a cold switch costs and the more \sys{} saves (\Cref{fig:where}c).
\full cuts switch-call latency by \nSensAFullFirst\% for contexts below 2K tokens and by \nSensAFullLast\% beyond 64K, and \oraclehost by up to \nSensAOracleHostLast\%.
Without the host tier, \harvestslo falls back above 32K, because long shadows are evicted from HBM.

\myparagraph{SLO strictness}
Tighter SLOs leave less slack to harvest (\Cref{fig:where}d; d--g average one pair per trace).
At half the default SLOs, with provisioning fixed, \full still cuts LLM-call latency by \nSensBFullFirst\%; at the default it cuts \nSensBFullDefault\%, and at 2--3$\times$ looser SLOs about \nSensBFullLast\%.

\myparagraph{Load}
As normal traffic grows, idle capacity shrinks and \harvestslo's gain falls from \nSensCSloFirst\% at $\rho{=}0.2$ to \nSensCSloLast\% at $\rho{=}0.9$ (\Cref{fig:where}e).
At the same time, cold switch calls queue longer and pools evict more, so the host tier matters more and \full's gain grows to \nSensCFullLast\%.

\myparagraph{Shadow KV cap and cold-entry age}
The shadow class rarely fills its 25\% cap: shadow KV averages \nKnobShadowOcc\% of the pool, because shadows enter only free memory and cold entries and are evicted first.
Raising the cap to the whole pool leaves \full at \nKnobCapFullFull\% (\Cref{fig:where}f); shrinking it to 5\% lowers \full only to \nKnobCapFullFive\%.
The cold-entry age sets which cached entries a shadow may displace (\Cref{fig:where}g).
At 0\,s, shadows displace entries that are about to be reused, including other shadows: shadow work grows \nKnobAgeZeroTok$\times$ and \full drops to \nKnobAgeFullZero\%.
At 120\,s, shadows fit only into free memory, the switch-call hit rate falls from \nKnobAgeHitDefault\% to \nKnobAgeHitLong\%, and \full drops to \nKnobAgeFullLong\%.
Any age between 2 and 10\,s gives \nKnobAgeFullTwo--\nKnobAgeFullTen\%.

\Cref{app:sensitivity} further varies the router, the controller parameters, the host tier, the cluster scale, the normal-traffic trace, the P/D configuration, and the chunk sizes in sensitivity study.

\providecommand{\testbedfigdir}{figures/gpu}
\newcommand{\nTestbedTtft}{61.9}
\newcommand{\nTestbedLatency}{19.8}
\newcommand{\nTestbedTask}{7.0}
\newcommand{\nTestbedCost}{24.6}
\newcommand{\nTestbedHitNone}{89.7}
\newcommand{\nTestbedHitShadow}{94.1}
\newcommand{\nTestbedRouterCost}{20.4}
\newcommand{\nTestbedEnergy}{1.4}

\section{Testbed Validation}
\label{sec:testbed}
We complement the simulation study in \Cref{sec:eval} with a vLLM-based prototype~\cite{kwon2023vllm}, evaluating \sys{} (\harvest) on a 16-H100 GPU testbed.

\myparagraph{Setup}
We evaluate two FusionBench~\cite{feng2025fusionbench} model pairs that are also used in simulation: Qwen2.5-7B and Llama-3.1-8B, each paired with Llama-3.1-70B.
To additionally evaluate learned routing and task accuracy, we use SWE-bench Verified~\cite{jimenez2024swebench}, since the GHCP-coding trace provides token lengths but not the prompt text needed for these experiments.
We pair SWE-agent-LM-7B with SWE-agent-LM-32B, both fine-tuned for software-engineering tasks, and include SWE-agent-LM-7B paired with Qwen2.5-72B-Instruct as a second configuration.
We measure task accuracy on a fixed random sample of 50 tasks from the 500-task SWE-bench Verified set, then replay recorded agent calls and routing decisions for serving-performance comparisons.
\Cref{app:testbed-router,app:testbed-accuracy} provide the routing and accuracy-evaluation details.
To evaluate serving performance separately, we replay recorded calls and routing decisions with and without shadow prefill (\harvest).

Each model instance also serves normal traffic with Poisson arrivals and Azure-derived request lengths, following the methodology in \Cref{sec:eval}.
We set the normal-traffic arrival rate to $\rho{=}0.3$ of each instance.
\Cref{app:testbed-setup} specifies the measurement windows, call matching, and uncertainty estimates.

\begin{figure}[t!]
  \centering
  \includegraphics[scale=0.95]{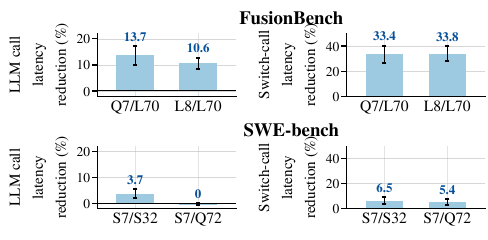}
  \vspace{-4mm}
  \caption{LLM-call and switch-call latency reductions.}
  \label{fig:testbed-latency}
  \vspace{-3mm}
\end{figure}

\myparagraph{LLM time per task}
\Cref{fig:testbed} shows that \sys{} reduces switch-call TTFT by \nTestbedTtft\% on average through prefix reuse.
Both FusionBench pairs also achieve lower LLM-call latency, consistent with simulation (\Cref{fig:testbed-latency}).
The reduction in LLM time per task is smaller because shadow prefill accelerates switch-call prefill, not decoding. It averages 12.1\% on FusionBench and \nTestbedTask\% across all pairs.
S7/Q72 illustrates this distinction: only 13\% of calls switch models, and decoding dominates execution. A 65.9\% switch-call TTFT reduction therefore leaves its LLM time per task approximately unchanged (\Cref{tab:testbed-workload}).

\begin{figure}[t!]
  \centering
  \includegraphics[scale=0.97]{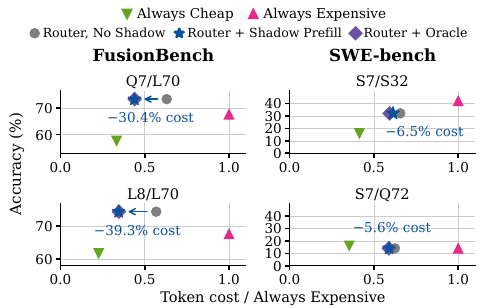}
  \vspace{-4mm}
  \caption{Accuracy versus normalized token cost. SWE-bench shadow costs use replay-measured cache reuse.}
  \label{fig:testbed-accuracy}
  \vspace{-3mm}
\end{figure}

\myparagraph{Accuracy and token cost}
\Cref{fig:testbed-accuracy} shows that \sys{} lowers estimated token cost by \nTestbedRouterCost\% on average.
The cost reduction holds on both benchmarks, consistent with simulation, even when decoding limits task-time gains.
Accuracy, in contrast, depends on the selected model pair.
On SWE-bench, routing with the agent-tuned S32 resolves 32\% of tasks, compared with 14\% using the larger, general instruction-tuned Q72.
Thus, larger model sizes alone do not ensure higher agent accuracy (more details in \Cref{app:testbed-accuracy}).

\begin{figure}[t!]
  \centering
  \includegraphics[scale=0.95]{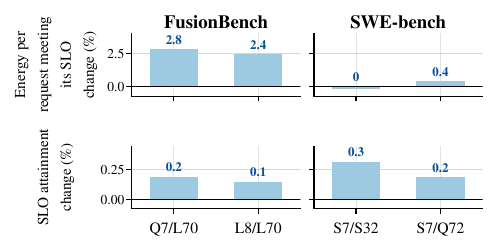}
  \vspace{-4mm}
  \caption{Energy and traffic SLO changes relative to \nosh.}
  \label{fig:testbed-overheads}
  \vspace{-3mm}
\end{figure}

\myparagraph{Energy and normal traffic}
\Cref{fig:testbed-overheads} shows a \nTestbedEnergy\% average increase in energy per SLO-compliant request relative to \nosh. Normal-traffic SLO attainment improves slightly on every pair.
Thus, at the evaluated load, the TTFT and cost benefits require little additional energy without compromising normal-traffic SLO attainment.

\section{Related Work}
\label{sec:related}

\myparagraph{Prefix caching and KV management}
vLLM, SGLang, and CachedAttention reuse KV across requests that share a prefix~\cite{kwon2023vllm,zheng2024sglang,gao2024cachedattention}; Mooncake, Pensieve, and LMCache extend the cache to host memory and remote storage~\cite{qin2024mooncake,yu2025pensieve,cheng2025lmcache}.
InferCept, KVFlow, and Continuum keep or prefetch a program's KV across tool calls~\cite{abhyankar2024infercept,pan2025kvflow,continuum2025}.
\textit{All of them manage the cache of the model that produced it; \sys{} builds a cache on a model that has not served the session.}
Across models, recent works translate one model's KV into another's~\cite{liu2024droidspeak,fu2026cache2cache,latentalign2026,heo2026crossmodel}, and ICaRus and PrefillShare train models to share a cache~\cite{icarus2026,prefillshare2026}.
\textit{They require retraining and approximate the target's KV while \sys{} computes the exact KV (lossless).}

\myparagraph{Harvesting and co-location}
Harvest VMs and GPU co-location systems run low-priority work in spare capacity~\cite{ambati2020harvestvm,han2022reef,strati2024orion}.
ConServe co-serves offline batch jobs with online inference~\cite{qiao2024conserve}.
Sarathi-Serve bounds iteration time with chunked prefill~\cite{agrawal2024sarathi}.
\textit{\sys{} applies LLM inference harvesting uniquely to a new kind of work in agentic model routing.}

\section{Conclusion}
\label{sec:conclusion}

We characterized the switch tax in per-step LLM routing and presented \sys{}, a system that exploits idle GPU capacity to warm prefix caches on routed-to models without additional GPUs or compromising normal-traffic SLOs.
Across 24 model pairs and five agentic benchmarks and traces, \sys{} reduces switch-call latency by \nTopSwFullMin--\nTopSwFullMax\% and routed token cost by \nTopCostFullMin--\nTopCostFullMax\%.

\FloatBarrier
\bibliographystyle{mlsys2025}
\bibliography{references}

\section*{AI Usage Disclosure}
Generative AI tools, including ChatGPT, Claude, and GitHub Copilot, were used during this work to support code development and debugging, literature review, and manuscript preparation.
Those AI tools were primarily used to assisting with implementation, polishing the language, and making the presentation clearer.
The authors formulated the research questions, designed the system and experimental methodology, and interpreted and validated the results.
The authors also reviewed and verified all AI-assisted code and written content.

\clearpage
\appendix
\begin{figure*}[t!]
  \centering
  {\makeatletter\def\@captype{table}\makeatother
  \caption{The 22 simulated model pairs; \Cref{tab:testbed-labels} lists the two SWE-bench pairs of the GPU testbed. Small and large are by total parameters; Params gives total (active) parameters for MoE models. Label is the pair's name in the figures used in the main paper (e.g., \Cref{fig:ttft}).}
  \label{tab:pairs}
  \small
  \setlength{\tabcolsep}{4pt}
  \begin{tabular}{@{}lllll@{}}
\toprule
Trace & Small model & Large model & Params (B) & Label in figures \\
\midrule
xRouteBench & GPT-OSS-20B & GPT-OSS-120B & 20.9 (3.6) / 116.8 (5.1) & OSS-20B / OSS-120B \\
 & GPT-OSS-20B & Llama-3.3-70B-Instruct & 20.9 (3.6) / 70.6 & OSS-20B / L3.3-70B \\
 & Llama-3.3-70B-Instruct & GPT-OSS-120B & 70.6 / 116.8 (5.1) & L3.3-70B / OSS-120B \\
 & Qwen2.5-7B-Instruct & Llama-3.3-70B-Instruct & 7.6 / 70.6 & Q2.5-7B / L3.3-70B \\
 & Qwen2.5-7B-Instruct & Qwen3-Next-80B-A3B-Instruct & 7.6 / 81.3 (3.8) & Q2.5-7B / Q3N-80B \\
\midrule
FusionBench & Llama-3.1-8B-Instruct & Llama-3.1-70B-Instruct & 8.0 / 70.6 & L3.1-8B / L3.1-70B \\
 & Llama-3.1-8B-Instruct & Mistral-Nemo-12B-Instruct & 8.0 / 12.2 & L3.1-8B / Nemo-12B \\
 & Mistral-7B-Instruct-v0.3 & Llama-3.1-70B-Instruct & 7.2 / 70.6 & M-7B / L3.1-70B \\
 & Mistral-7B-Instruct-v0.3 & Mixtral-8x22B-Instruct-v0.1 & 7.2 / 140.6 (39.1) & M-7B / Mx-8x22B \\
 & Qwen2.5-7B-Instruct & Llama-3.1-70B-Instruct & 7.6 / 70.6 & Q2.5-7B / L3.1-70B \\
\midrule
Router-R1 & Llama-3.1-70B-Instruct & Mixtral-8x22B-Instruct-v0.1 & 70.6 / 140.6 (39.1) & L3.1-70B / Mx-8x22B \\
 & Llama-3.1-8B-Instruct & Llama-3.1-70B-Instruct & 8.0 / 70.6 & L3.1-8B / L3.1-70B \\
 & Mistral-7B-Instruct-v0.3 & Gemma-2-27B-IT & 7.2 / 27.2 & M-7B / G2-27B \\
 & Mistral-7B-Instruct-v0.3 & Llama-3.1-70B-Instruct & 7.2 / 70.6 & M-7B / L3.1-70B \\
 & Qwen2.5-7B-Instruct & Llama-3.1-70B-Instruct & 7.6 / 70.6 & Q2.5-7B / L3.1-70B \\
 & Qwen2.5-7B-Instruct & Mixtral-8x22B-Instruct-v0.1 & 7.6 / 140.6 (39.1) & Q2.5-7B / Mx-8x22B \\
\midrule
GHCP-coding & Gemma-2-9B-IT & Granite-34B-Code-Instruct & 9.2 / 33.7 & G2-9B / Granite-34B \\
 & Gemma-2-9B-IT & Mixtral-8x7B-Instruct-v0.1 & 9.2 / 46.7 (12.9) & G2-9B / Mx-8x7B \\
 & Gemma-7B & Mixtral-8x7B-Instruct-v0.1 & 8.5 / 46.7 (12.9) & G-7B / Mx-8x7B \\
 & Llama-3.1-8B-Instruct & Llama-3.1-70B-Instruct & 8.0 / 70.6 & L3.1-8B / L3.1-70B \\
 & Mistral-7B-Instruct-v0.3 & Mixtral-8x22B-Instruct-v0.1 & 7.2 / 140.6 (39.1) & M-7B / Mx-8x22B \\
 & Qwen2.5-7B-Instruct & Llama-3.1-70B-Instruct & 7.6 / 70.6 & Q2.5-7B / L3.1-70B \\
\bottomrule
\end{tabular}
\par}
  \vspace{2mm}
  \includegraphics[width=0.83\textwidth]{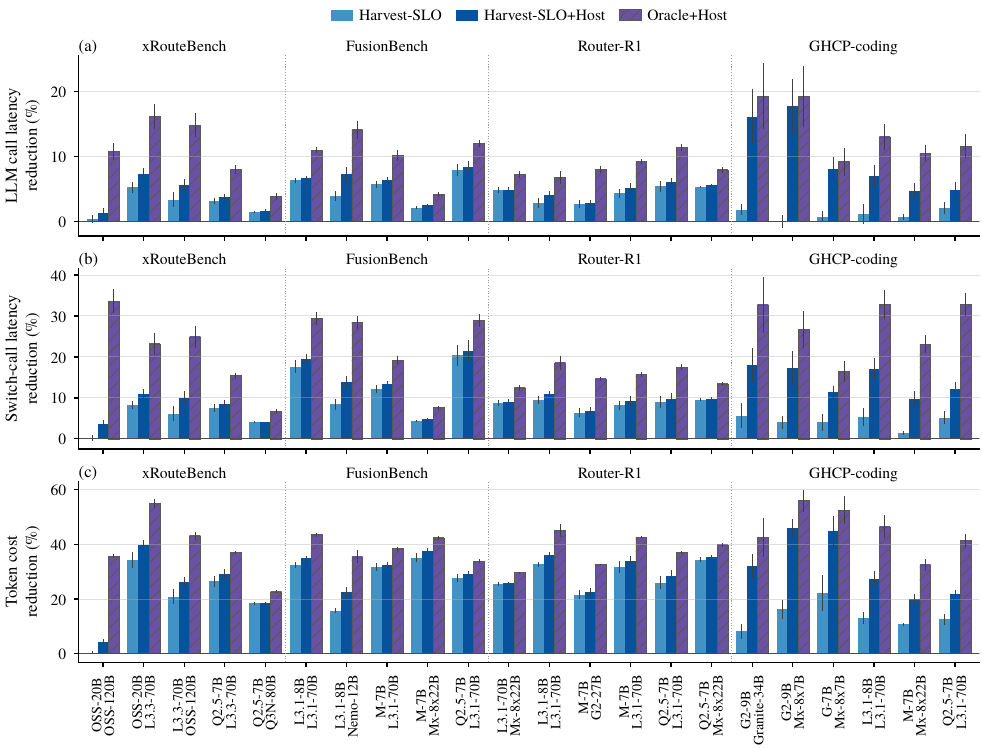}
  \vspace{-6mm}
  \caption{(a) LLM-call latency, (b) switch-call latency, and (c) routed token-cost reduction relative to \nosh for all 22 simulated model pairs. Error bars: $\pm$1 SE over 3 seeds $\times$ 3 windows.}
  \label{fig:allpairs}
\end{figure*}

\section{Results for All Model Pairs}
\label{app:allpairs}

\Cref{tab:pairs} lists the model pairs with their full names, and \Cref{fig:allpairs} shows the main metrics for every pair.
Over all 22 simulated pairs, \full cuts LLM-call latency by \nAllLatFullMin--\nAllLatFullMax\% (mean \nAllLatFull\%), switch-call latency by \nAllSwFullMin--\nAllSwFullMax\% (mean \nAllSwFull\%), switch-call TTFT by \nAllTtftFullMin--\nAllTtftFullMax\% (mean \nAllTtftFull\%), and routed token cost by \nAllCostFullMin--\nAllCostFullMax\% (mean \nAllCostFull\%).
Every pair gains in all four metrics.

\section{Simulation Details}
\label{app:sim}

\myparagraph{Deployments}
Each model runs on the smallest tensor-parallel degree at which its weights take at most half of HBM: 7--12B models and GPT-OSS-20B on one H100; Gemma-2-27B, GPT-OSS-120B, and Granite-34B on two; Llama-3.x-70B and Mixtral-8x7B on four; Mixtral-8x22B and Qwen3-Next-80B-A3B on eight.

\myparagraph{Latency model in Vidur}
An iteration contains prefill chunks ($q$ new tokens on $c$ cached tokens) and decodes.
Per layer, GEMMs take the maximum of compute and weight-read time; prefill attention adds compute and KV reads; decode attention adds KV reads and a per-sequence cost; tensor parallelism adds two all-reduces.
MoE layers use active parameters for compute and total parameters for weight reads~\cite{agrawal2024vidur}.
All operator latencies are grounded in measurements on H100 GPUs, providing high-fidelity estimates of prefill and decode execution times.

\myparagraph{Disaggregated cluster}
Each model's instances split into prefill and decode instances; the split (mostly 3P:1D) is the one with the highest SLO-compliant capacity for each model and normal-traffic trace.
A prefill instance runs prefill-only iterations; KV moves to the decode instance, and the transfer and decode-queue wait count toward TPOT.
A decode instance keeps a session's context after each call, so the next transfer sends only the missing tokens.
Load is set to $\rho{=}0.7$ of each cluster's own capacity, so colocated and disaggregated runs compare at the same relative load.

\section{Additional Sensitivity Studies}
\label{app:sensitivity}

\begin{figure*}[t]
  \centering
  \includegraphics[width=\textwidth]{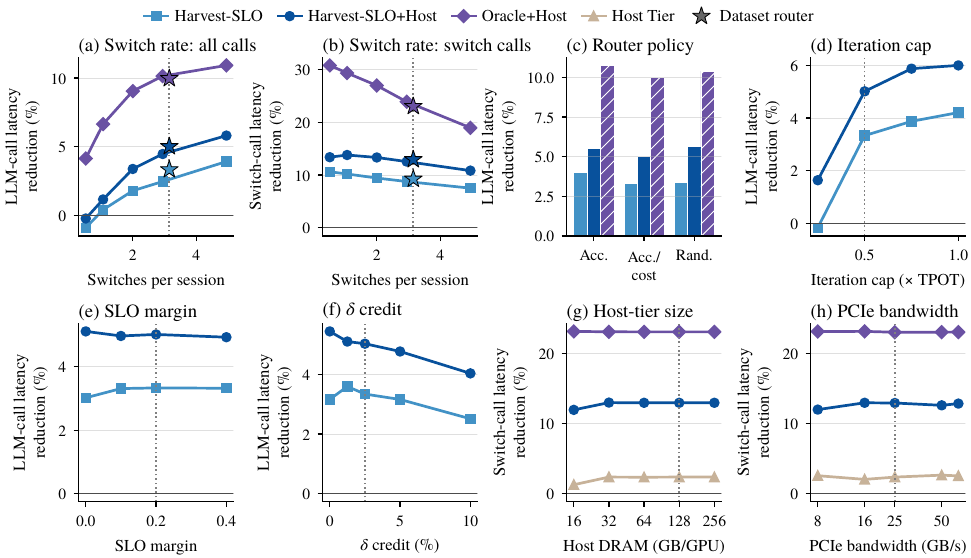}
  \vspace{-6mm}
  \caption{Sensitivity to the router and the controller. (a) LLM-call and (b) switch-call latency reduction versus switches per session under a Markov router (stars: dataset router). LLM-call latency reduction under (c) the max-accuracy, max-accuracy-per-cost (default), and random routing policies, and versus (d) the iteration cap $\beta$, (e) the SLO margin $m$, and (f) the credit rate $\delta$. Switch-call latency reduction versus (g) host-tier size and (h) PCIe bandwidth. Mean over the four setups of \Cref{fig:where}d--g; dotted lines mark the defaults.}
  \label{fig:sens-workload}
  \vspace{-3mm}
\end{figure*}

\begin{figure*}[t]
  \centering
  \includegraphics[width=\textwidth]{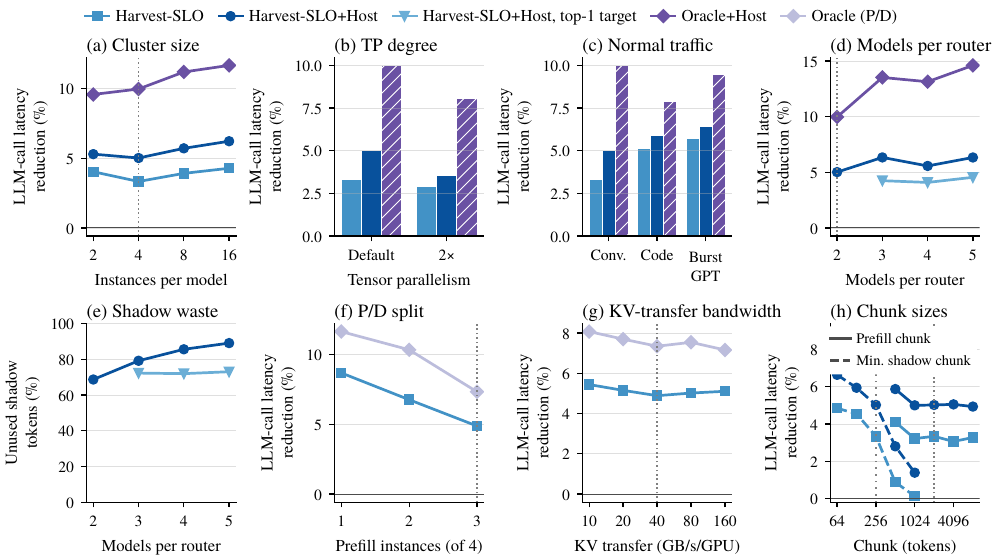}
  \vspace{-6mm}
  \caption{Sensitivity to the deployment. LLM-call latency reduction versus (a) instances per model, (b) tensor-parallel degree (default: \Cref{app:sim}; 2$\times$: twice that degree), (c) normal-traffic trace, and (d) models per router; (e) shadow tokens never used by a later call. In (d, e), \full shadows the session to every other model; the top-1 variant shadows only to the most likely next model. P/D-disaggregated cluster: LLM-call latency reduction versus (f) prefill instances out of four per model and (g) KV-transfer bandwidth. (h) Prefill chunk budget (solid) and minimum shadow chunk (dashed). Mean over the four setups of \Cref{fig:where}d--g; dotted lines mark the defaults.}
  \label{fig:sens-deploy}
  \vspace{-3mm}
\end{figure*}

We vary one parameter at a time on the four setups of \Cref{fig:where}d--g (one pair per trace), with 3 seeds $\times$ 3 windows at $\rho{=}0.7$, and report the mean over the setups.
Reductions are relative to \nosh on the calls that complete under every compared method.
When a parameter changes the environment \nosh runs in (router, cluster, normal traffic, prefill chunk, P/D configuration, models per router), \nosh is evaluated at the same setup for a fair comparison.
Cluster size and tensor parallelism use capacities measured for that configuration, and load stays at $\rho{=}0.7$ of that capacity, so agentic and normal traffic scale with the cluster.
At the defaults, \harvestslo and \full reduce LLM-call latency by \nKnobCapSloDefault\% and \nKnobCapFullDefault\%, as in \Cref{fig:where}.

\myparagraph{Router switch rate and routing policy}
A two-state Markov router replaces the dataset's choices and switches model at each step with probability $p \in [0.05, 0.5]$, giving \nAppSwSessLow{} to \nAppSwSessHigh{} switches per session; the dataset routers give \nAppSwSessData.
More switches give shadows more calls to help: \full's LLM-call reduction grows from \nAppRtFullLow\% to \nAppRtFullHigh\% and \oraclehost's from \nAppRtOrLow\% to \nAppRtOrHigh\% (\Cref{fig:sens-workload}a).
The gain per switch call falls, from \nAppRtSwFullLow\% to \nAppRtSwFullHigh\% for \full and from \nAppRtSwOrLow\% to \nAppRtSwOrHigh\% for \oraclehost (\Cref{fig:sens-workload}b): with frequent switches, the session left the target model only a few steps earlier, so \nosh already finds more of the context cached.
Re-routing the sessions with the accuracy-maximizing or random policy changes the switch rate (\nAppPolSwSessRand{} switches per session under random routing), and \full reduces LLM-call latency by \nAppPolFullAcc\%, \nAppPolFullCost\% (default), and \nAppPolFullRand\% (\Cref{fig:sens-workload}c).

\myparagraph{Controller parameters}
Of the three controller parameters, the iteration cap $\beta$ matters most.
At $\beta{=}0.25$ TPOT, few shadow chunks fit and \harvestslo gives \nAppCapSloQ\% (\full \nAppCapFullQ\%); at $\beta{=}1$, \harvestslo gives \nAppCapSloOne\% and \full \nAppCapFullOne\% (\Cref{fig:sens-workload}d).
The SLO margin $m$ moves \full only within \nAppMarginFullMin--\nAppMarginFullMax\% (\Cref{fig:sens-workload}e): most in-flight requests are far ahead of their SLO, so $\beta$, not the slack, bounds the budget.
A larger credit rate $\delta$ lets shadow chunks stretch iterations that carry routed calls: \full falls from \nAppDeltaFullZero\% at $\delta{=}0$ to \nAppDeltaFullTen\% at $\delta{=}0.1$, and \harvestslo peaks at \nAppDeltaSloMax\% (\Cref{fig:sens-workload}f).
Across all controller settings, normal-traffic SLO attainment differs from \nosh by only \nAppCtrlAttMin{} to \nAppCtrlAttMax{} percentage points.

\myparagraph{Host tier}
The host tier needs only 32\,GB per GPU: at 16\,GB, \full's switch-call reduction drops from \nAppHostSwFullDefault\% to \nAppHostSwFullSmall\% and \hosttier's from \nAppHostSwTierDefault\% to \nAppHostSwTierSmall\%, while 64, 128, and 256\,GB give identical results (\Cref{fig:sens-workload}g).
PCIe bandwidth from 8 to 64\,GB/s moves \full only within \nAppBwSwFullMin--\nAppBwSwFullMax\% (\Cref{fig:sens-workload}h); \oraclehost stays at \nAppHostSwOr\% throughout (at no switch cost regardless of PCIe).

\myparagraph{Cluster scale}
\full reduces LLM-call latency by \nAppInstFullTwo\% with 2 instances per model and \nAppInstFullSixteen\% with 16, and \oraclehost by \nAppInstOrTwo\% and \nAppInstOrSixteen\% (\Cref{fig:sens-deploy}a).
Doubling each model's tensor-parallel degree shortens prefill, so a cold switch call costs less and every method gains less: \full \nAppTpFull\%, \harvestslo \nAppTpSlo\%, and \oraclehost \nAppTpOr\% (versus \nAppTpOrDefault\%; \Cref{fig:sens-deploy}b).

\myparagraph{Normal traffic}
With code or BurstGPT normal traffic instead of conversations, \harvestslo reduces LLM-call latency by \nAppBgSloCode\% and \nAppBgSloBurst\% (versus \nAppBgSloConv\%), and \full by \nAppBgFullCode\% and \nAppBgFullBurst\% (\Cref{fig:sens-deploy}c).
Normal-traffic SLO attainment rises by at most \nAppBgSloAttBurst{} percentage points.

\myparagraph{More than two models per router}
Each setup's pair grows along its trace's candidate models to $N{=}3$--5, and the sessions are re-routed by the accuracy-per-cost policy over all $N$ models, raising switches per session from \nAppSwSessData{} to \nAppNmSwSessFive.
With $N{>}2$, the next model of a switch is uncertain.
\full shadows the session to all other models, ordered by a leave-one-session-out prior on the next model; the top-1 variant shadows only to the most likely next model.
Shadowing all targets reduces LLM-call latency by \nAppNmAllThree\% at $N{=}3$ and \nAppNmAllFive\% at $N{=}5$, versus \nAppNmTopThree\% and \nAppNmTopFive\% for top-1 (\Cref{fig:sens-deploy}d).
The switch-call reduction of \full falls from \nAppNmSwAllTwo\% at $N{=}2$ to \nAppNmSwAllFive\% at $N{=}5$, and the share of its shadow tokens that go unused grows from \nAppNmWasteAllTwo\% to \nAppNmWasteAllFive\% (top-1 at $N{=}5$: \nAppNmWasteTopFive\%; \Cref{fig:sens-deploy}e); energy per request rises by \nAppNmEnergyAllFive\% (top-1: \nAppNmEnergyTopFive\%).
\oraclehost grows from \nAppNmOrTwo\% to \nAppNmOrFive\%, so better next-model prediction would recover more of the benefit from shadow prefill.

\myparagraph{P/D disaggregation}
With fewer prefill instances, prefill queues longer and a cold switch call costs more: \harvestslo reduces LLM-call latency by \nAppPdSloOne\% with one prefill instance out of four and \nAppPdSloThree\% with three (default), and \oracle by \nAppPdOrOne\% and \nAppPdOrThree\% (\Cref{fig:sens-deploy}f).
KV-transfer bandwidth from 10 to 160\,GB/s per GPU moves \harvestslo only within \nAppXferSloMin--\nAppXferSloMax\% (\Cref{fig:sens-deploy}g).

\myparagraph{Chunk sizes}
The prefill chunk budget matters little: \full gives \nAppChunkBigFullMin--\nAppChunkBigFullMax\% from 1,024 to 8,192 tokens and \nAppChunkFullSmall\% at 512 (\Cref{fig:sens-deploy}h).
Shadow chunks below a minimum size (256 tokens) are deferred, since each chunk re-reads the whole cached context.
This minimum matters more: at 64 tokens, \full reaches \nAppMinFullSixtyFour\%; at 512 and 1,024 tokens, chunks rarely fit in the budget and \full drops to \nAppMinFullFiveTwelve\% and \nAppMinFullKilo\%.

\providecommand{\testbedfigdir}{figures/gpu}

\section{Testbed Details}
\label{app:testbed}

\subsection{Serving Protocol}
\label{app:testbed-setup}

\begin{table*}[t]
  \centering
  \caption{Testbed deployments and figure labels. Parentheses give H100-80GB GPU count and tensor-parallel degree (TP) per model instance.}
  \label{tab:testbed-labels}
  \par\medskip
  \small
  \setlength{\tabcolsep}{4pt}
  \renewcommand{\arraystretch}{1.1}
  \begin{tabularx}{\textwidth}{@{}lXXl@{}}
    \toprule
    Benchmark & Small model & Large model & Figure label \\
    \midrule
    FusionBench & Qwen2.5-7B-Instruct\newline (1 GPU, TP=1) & Llama-3.1-70B-Instruct\newline (4 GPUs, TP=4) & Q7/L70 \\
     & Llama-3.1-8B-Instruct\newline (1 GPU, TP=1) & Llama-3.1-70B-Instruct\newline (4 GPUs, TP=4) & L8/L70 \\
    \midrule
    SWE-bench & SWE-agent-LM-7B\newline (1 GPU, TP=1) & SWE-agent-LM-32B\newline (2 GPUs, TP=2) & S7/S32 \\
     & SWE-agent-LM-7B\newline (1 GPU, TP=1) & Qwen2.5-72B-Instruct\newline (4 GPUs, TP=4) & S7/Q72 \\
    \bottomrule
  \end{tabularx}
\end{table*}

\myparagraph{Prototype}
Each model uses one vLLM instance with prefix caching, chunked prefill, and priority scheduling (\Cref{tab:testbed-labels}).
After a call completes, the manager submits its prompt and response to the other model as lowest-priority shadow work with one output token.
For non-append-only histories, it uses the next prompt's prefix, capped at the completed prompt length.
The destination computes its own KV, without cross-model transfer.
Newer completed calls supersede pending shadows.
Admission requires an empty destination waiting queue and permits one in-flight shadow per instance.
KV-usage and SLO-based admission do not constrain these runs.
The prototype omits per-iteration credit, the SLO-slack controller, evict-first shadow KV, and the host tier.

\myparagraph{Traffic and replay}
Normal traffic follows Poisson arrivals at $\rho$ times each instance's calibrated capacity.
Capacity is the highest tested rate with at least 90\% attainment of both TTFT and TPOT SLOs.
Their bounds are derived as $\max(1\,\mathrm{s},5t_{\mathrm{prefill}}(p))$ and $\max(25\,\mathrm{ms},5t_{\mathrm{decode}})$, using isolated prefill and batch-1 decode latencies.
Capacities are 12.8, 11.0, 3.8, 13.3, 4.8, and 3.8 requests/s for Qwen2.5-7B, Llama-3.1-8B, Llama-3.1-70B, SWE-agent-LM-7B, SWE-agent-LM-32B, and Qwen2.5-72B, respectively.
Azure traces~\cite{stojkovic2025dynamollm} supply input and output lengths capped at 8,192 and 1,024 tokens.
FusionBench and SWE-bench session-arrival rates are 0.05/s and 0.01/s.
SWE-bench samples with replacement from 30 S7/S32 and 95 S7/Q72 trajectories.
Replay fixes routes, prompt structure, output lengths, and tool observations.
Each call follows predecessor completion plus a lognormal tool gap with median 2\,s.
Append-only histories include replay-generated responses.

\myparagraph{Measurement}
FusionBench and SWE-bench runs last 900\,s and 1,320\,s, excluding the first 120\,s.
We match calls by session and step.
Cache hit rate is cached input divided by total input tokens.
LLM time per task excludes tool execution and uses the same matched latency sums as the all-call reduction. NVML measures total serving-GPU energy, including idle power and normal traffic, divided by the number of SLO-compliant requests.

\begin{table*}[t]
  \centering
  \caption{Replay workload under \nosh at $\rho{=}0.3$. Lengths are tokens per call.}
  \label{tab:testbed-workload}
  \par\medskip
  \small
  \setlength{\tabcolsep}{4pt}
  \renewcommand{\arraystretch}{1.1}
  \begin{tabular*}{\textwidth}{@{\extracolsep{\fill}}lrrrrrr@{}}
    \toprule
    Model pair & \shortstack{Switch\\calls} & \shortstack{Switches\\per session} & \shortstack{Mean\\prompt} & \shortstack{p99\\prompt} & \shortstack{Mean\\output} & \shortstack{Hit rate\\(\%)} \\
    \midrule
    Q7/L70 & 20\% & 2.5 & 19K & 75K & 52 & 85.2 \\
    L8/L70 & 26\% & 3.2 & 19K & 75K & 97 & 85.0 \\
    \midrule
    S7/S32 & 43\% & 15.4 & 41K & 93K & 175 & 93.9 \\
    S7/Q72 & 13\% & 6.3 & 33K & 68K & 249 & 96.6 \\
    \bottomrule
  \end{tabular*}
\end{table*}

\myparagraph{Workload dependence}
\Cref{tab:testbed-workload} shows longer outputs for SWE-bench, which limit the task-time contribution of prefill savings.
High baseline reuse does not eliminate switch-call prefill: SWE-bench switch calls cache 85.8\% of input tokens on average, yet have 2.6$\times$ the same-model mean TTFT.

\begin{figure}[t]
  \centering
  \includegraphics[width=\columnwidth]{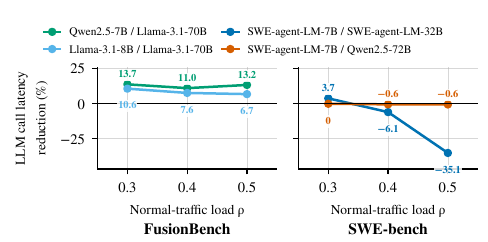}
  \vspace{-2mm}
  \caption{LLM-call latency reduction versus normal-traffic load at fixed session-arrival rates.}
  \label{fig:testbed-load}
  \vspace{-3mm}
\end{figure}

\myparagraph{Admission under load}
\Cref{fig:testbed-load} isolates the limit of queue-based admission.
At $\rho{=}0.5$, both FusionBench pairs retain latency reductions, however S7/S32 slows by 35.1\%.
This is becasue an empty waiting queue can coexist with running decodes, allowing shadow work to extend their iterations.
This motivates the iteration-level interference budget in \Cref{sec:design:harvest}.

\subsection{Router Evaluation}
\label{app:testbed-router}

\myparagraph{Training}
We train pair-specific bilinear matrix-factorization routers~\cite{llmrouter2025} on large-model trajectories from 50 tasks.
Labels measure model agreement, not task success.
S7/S32 uses teacher-forced greedy-token agreement thresholded at the training median.
S7/Q72 uses agreement on tool calls and key arguments.
Features combine embeddings of the last 4K prompt tokens, the latest action and observation, and compact history with standardized step features.
Three-fold task-level cross-validation yields AUCs of 0.76 and 0.74, respectively.
Training-cohort tasks use held-out-fold predictions, while other tasks use the fold ensemble.

\begin{figure*}[t]
  \centering
  \includegraphics[scale=0.9]{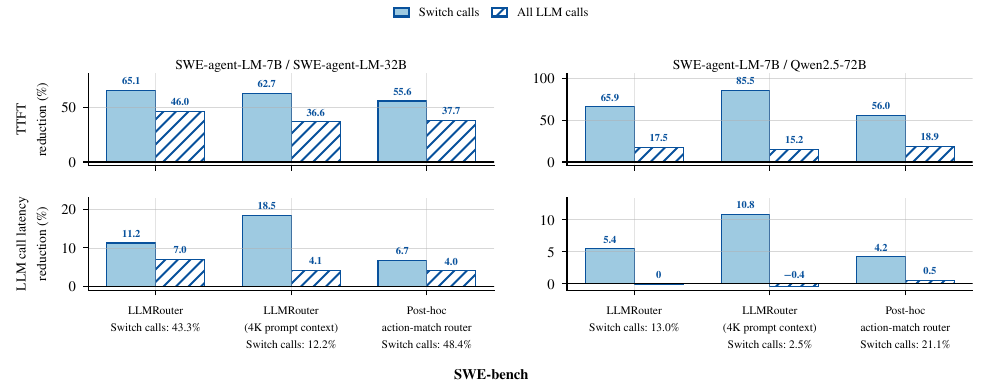}
  \vspace{-4mm}
  \caption{Shadow-prefill latency reductions relative to each router's own \nosh baseline. Solid bars show switch calls. Hatched bars show all calls.}
  \label{fig:testbed-router}
  \vspace{-3mm}
\end{figure*}

\myparagraph{Sensitivity}
\Cref{fig:testbed-router} compares the default router, a last-4K-prompt-only variant, and offline action-match routing on 15 tasks per pair at $\rho{=}0.3$.
Action-match chooses the small model when recorded actions agree.
It requires both models' actions and is not deployable.
Each policy's shadow and \nosh calls are matched, so bars compare serving benefits rather than accuracy or absolute router latency.
For S7/Q72, prompt-only routing improves switch-call TTFT more than the default (85.5\% versus 65.9\%), but switches less often (2.5\% versus 13.0\%).
Its all-call TTFT reduction is therefore smaller (15.2\% versus 17.5\%).
Switch frequency and per-switch benefit jointly determine workload-wide improvement.

\subsection{Accuracy}
\label{app:testbed-accuracy}

\begin{table}[t]
  \centering
  \caption{Resolved SWE-bench Verified tasks on the fixed 50-task cohort.}
  \label{tab:testbed-accuracy}
  \par\medskip
  \small
  \setlength{\tabcolsep}{3pt}
  \begin{tabular*}{\columnwidth}{@{\extracolsep{\fill}}lr@{}}
    \toprule
    Configuration & Accuracy (\%) \\
    \midrule
    SWE-agent-LM-7B & 16.0 \\
    SWE-agent-LM-32B & 42.0 \\
    Qwen2.5-72B-Instruct & 14.0 \\
    \midrule
    S7/S32 router & 32.0 \\
    S7/Q72 router & 14.0 \\
    \bottomrule
  \end{tabular*}
\end{table}

\myparagraph{Agent protocol}
We execute SWE-agent with the SWE-smith tool configuration~\cite{yang2025swesmith} and grade all 50 tasks with the official harness (\Cref{tab:testbed-accuracy}).
Unappliable patches count as unresolved.
To preserve cumulative prefixes, we disable replacement of older tool observations with placeholders.
SWE-bench models use YaRN with a 131,072-token context limit, without sliding-window history or silent truncation.
Tool-output bounds are unchanged.
We remove the monetary stopping budget and retain the 75-call limit.

\section{Discussion}
\label{sec:discussion}

\myparagraph{More than two models for model switches}
A router can choose among any number of models.
With $N$ models, a session can switch to any of the other $N{-}1$, so \sys{} must also guess which model comes next.
\sys{} ranks the candidates by a prior on the next model and spends shadow capacity on the most likely target first (\Cref{sec:design:jobs}).
With five models, shadowing all targets still reduces LLM-call latency by \nAppNmAllFive\%, but \nAppNmWasteAllFive\% of shadow tokens go unused and energy per request rises by \nAppNmEnergyAllFive\%; shadowing only the top-1 target gives \nAppNmTopFive\% at \nAppNmEnergyTopFive\% more energy (\Cref{app:sensitivity}).
Meanwhile, \oraclehost grows from \nAppNmOrTwo\% to \nAppNmOrFive\%, because more models mean more switches.
The gap between the two comes from next-model prediction, which a router can close by exposing its scores for upcoming calls, or which an agent's plan can supply when it fixes the model of the next steps ahead of time.

\myparagraphnodot{What does it mean for the router?}
Switch-aware routers penalize switches~\cite{dialrouter2026,mtrouter2026}, because a switch has so far cost a cold prefill.
With \sys{}, the cost of a switch depends on how much of the candidate's prefix is warm, and the router can query that state and price each switch accordingly.
A router that switches more freely gains accuracy (\Cref{sec:eval:acc}), and \sys{} makes those switches cheaper: with frequent switches, \full reduces LLM-call latency by up to \nAppRtFullHigh\% (\Cref{app:sensitivity}).
Router and serving system can also cooperate in the other direction: the router's switch probabilities can drive the ordering of shadow jobs, and a more accurate predictor lets a gate such as \predictive skip unlikely targets while missing fewer switches.

\myparagraph{Finer-grained model switching}
\sys{} targets routers that decide per LLM call.
Systems that switch within a call, per token or per segment, such as speculative decoding~\cite{leviathan2023spec,chen2023specsampling}, speculative cascades~\cite{narasimhan2025speccascades}, and token-level routing~\cite{whiflash2026}, leave no gap between the decision and the next model's use of the context.
These systems keep both models' KV current at every step by construction, and pay for it on the critical path of decoding.
\sys{} instead trails the routed model by the output of the current call and runs only in harvested slack, which suits decisions made once per call.
Cross-model KV translation~\cite{liu2024droidspeak,fu2026cache2cache,heo2026crossmodel} could fill the remaining gap at a switch for models in the same family, at the cost of approximate rather than exact KV.

\myparagraph{When \sys{} helps less}
\sys{} saves prefill, so its gain shrinks when prefill is a small part of a switch call.
For targets above 50B, decoding dominates and switch-call latency falls by \nSizeBigLat\% (\Cref{sec:eval:where}); doubling the tensor-parallel degree shortens prefill and lowers \full's LLM-call gain to \nAppTpFull\% (\Cref{app:sensitivity}).
Long reasoning outputs, as from DeepSeek-R1, leave the target's instances little slack to harvest.
Rare switches leave little to warm: at about \nAppSwSessLow{} switches per session, \full's LLM-call gain is \nAppRtFullLow\%.
Under heavy normal traffic, the SLO-slack controller finds little capacity (\harvestslo gives \nSensCSloLast\% at $\rho{=}0.9$), and the host tier provides most of \full's gain.

\myparagraph{Wasted work and energy}
Shadow prefill is speculative: \nShadowWasteFull\% of \full's shadow tokens are never used, because the router decides only when the call arrives.
On colocated clusters this work rides on iterations that run anyway, and energy per good request rises by \nEnergyFull\%.
On P/D-disaggregated clusters, shadow work wakes otherwise-idle prefill GPUs and raises energy by \nPdEnergy\% (\Cref{sec:eval:overheads}).
Operators that cap power or bill energy can gate shadow work by switch probability, or limit it to the top-1 target, trading part of the latency gain for less waste.

\end{document}